\documentclass[]{aa}  
\usepackage{amsmath}
\usepackage{epsfig}
\usepackage{natbib}
\bibpunct{(}{)}{;}{a}{}{,}
\usepackage{graphicx}
\usepackage{amssymb}
\usepackage{mathrsfs}
\usepackage{array}
\usepackage{listings}
\usepackage{soul}

\def\m2s2{\hbox{\,m$^{2}$\,s$^{-2}$}} 

\def\P~{\hbox{$\tilde{P}$}}

\usepackage{xspace}

\newcommand{\ve}[1]{\boldsymbol{#1}} 
\newcommand*{\eg}{e.g.\@\xspace}

\begin{document}

\title{Vortices in stratified protoplanetary disks:} 
\subtitle{From baroclinic instability to vortex layers}

\author{ P. Barge \inst{1} \and S. Richard \inst{1,2,3} \and S. Le Diz\`es \inst{3} }
\institute{Aix Marseille Universit\'e, CNRS, LAM UMR 7326, F-13013 Marseille, France \label{inst1} 
\and Aix Marseille Universit\'e, CNRS, Centrale Marseille, IRPHE UMR 7342, F-13013 Marseille, France  \label{inst2}
\and Astronomy Unit, Queen Mary University of London, Mile End Road, London E1 4NS, UK  \label{inst3} 
\\ \email{pierre.barge@lam.fr, samuel.richard@qmul.ac.uk, ledizes@irphe.univ-mrs.fr}   }

\date{Received    / Accepted    }

\abstract
{Large-scale vortices could play a key role in the evolution of protoplanetary disks, particularly in the dead-zone where no turbulence associated with magnetic field is expected. Their possible formation by the subcritical baroclinic instability is a complex issue because of the vertical structure of the disk and the elliptical instability.}  {In 2D disks the baroclinic instability is studied as a function of the thermal transfer efficiency. In three-dimensional disks we explore the importance of radial and vertical stratification on the processes of vortex formation and amplification.}
{Numerical simulations are performed using a fully compressible hydrodynamical code based on a second-order finite volume method. We assume a perfect gas law in inviscid disk models in which heat transfer is due to either  relaxation or  diffusion.}  
{In 2D, the baroclinic instability with thermal relaxation  leads to the formation of large-scale vortices, which are unstable with respect to the elliptic instability. In the presence of heat diffusion, hollow vortices are formed which evolve into vortical structures with a turbulent core. In 3D, the disk stratification is found to be unstable in a finite layer which can include the mid-plane or not.  When the unstable layer contains the mid-plane, the 3D baroclinic instability with thermal relaxation is found to develop first in the unstable layer as in 2D, producing large-scale vortices. These vortices are then stretched out in the stable layer,  creating long-lived columnar  vortical structures extending through the width of the disk. They are also found to be the source of internal vortex layers that develop across the whole disk along baroclinic critical layer surfaces, and form new vortices in the upper region of the disk.}
{In 3D disks, vortices can survive for a very long time if the production of vorticity by the baroclinic amplification balances the destruction of vorticity by the elliptical instability. However, this possibility is strongly dependent on the disk properties. Such baroclinic vortices could play a significant role in the global disk evolution and in participating to the decoupling of solids from the gas component. They could also contribute to the formation of new out-of-plane vortices by a critical layer excitation mechanism.}
\keywords{ protoplanetary disks - vortices - hydrodynamics - vortices }

\titlerunning{Vortices in stratified disks }

\authorrunning{P. Barge et al. }
\maketitle


\section{Introduction}
\label{sec:intro}
Large-scale vortices in protoplanetary disks could play an important role in the evolution of gas and dust at the decoupling stage of planet formation. For example, they could participate in the transport of angular momentum, providing a natural mechanism for accretion onto the star; they could also stop the systematic drift of the solid particles thanks to the trapping mechanism identified by  \citet{Barge95}. This is the reason why the formation, stability, and evolution of such vortices in protoplanetary disks deserve special attention. Recent observations with high resolution in the millimeter and centimeter range, have recently raised a strong interest on dust trapping and vortices.
Vortex formation in protoplanetary disks has been studied a lot in the last ten years. Their formation by the Rossby-wave instability was addressed first \citep{Lovelace99, Li2000} and studied in detail in 2D and 3D disks \citep{Li2001, Meheut2012a, Meheut2012b, Lin2012, Lin2013, Richard2013}. Other
mechanisms (like the baroclinic instability) were also proposed for the formation of vortices \citep{Klahr2004, Petersen2007a, Petersen2007b, Lesur2010}.

The baroclinic instability is well known in geophysical fluid dynamics where it is responsible for the formation of cyclones in the atmosphere of the Earth \citep{Pedlosky87}. It originates in the systematic bending of the iso-pressure surfaces with respect to the iso-density surfaces, which leads to a source term in the evolution equation of the vorticity field, the so-called baroclinic source term. It is only recently that this instability was actually addressed in the context of protoplanetary disks.  \citet{Klahr2003} explored the possibility of finding a baroclinic instability in 2D protoplanetary disks with global numerical simulations. They reported the formation of large-scale vortices, but the actual baroclinic origin of these vortices is unclear since thermal transfer is absent from their simulations, contrary to what is normally required (\citet{Petersen2007a,Petersen2007b}). The non-linear nature of this instability was underlined by \citet{Lesur2010} (thereafter LP10) using incompressible shearing box simulations, clearly explaining why the works of \citet{Klahr2004}  and \citet{Johnson2005a} failed to describe the baroclinic amplification of the vortices.
In the 2D context, \citet{Raettig2013} have tried to estimate the generation of baroclinic turbulence in protoplanetary disks. They studied the dependence of vortex amplification as a function of the entropy gradient, the thermal transfer efficiency and the numerical resolution of the computations. 

\citet{Lyra2011} also investigated numerically the growth of the subcritical baroclinic instability in a magnetic disk; they concluded that baroclinic vortices are destroyed by the magneto-rotational instability and can only survive in the dead zone of the protoplanetary disks.

This instability was then studied by LP10 who noted that the baroclinic instability in protoplanetary disks is basically non-linear and differs from the linear instability that forms cyclones in geophysical flows. They also found that, in appropriate conditions, 3D baroclinic vortices may survive the elliptical instability. \citet{Klahr2014} recently proposed that a convective overstability could develop following a linear amplification of the epicyclic oscillations in axisymmetric and vertically unstratified disks and \citet{Lyra2014} showed using 3D simulations, it could form vortices. In this paper, this mechanism is not considered. Here, we focus on the subcritical baroclinic instability as described by \citet{Lesur2010}.

LP10 performed three dimensional simulations to study the stability of baroclinic vortices without vertical stratification and with a radial stratification only. However, to our knowledge, the baroclinic instability has never been studied in a fully stratified three dimensional disk. In this paper, we revisit the problem using fully compressible and long term numerical simulations. First, we start from numerical simulations in a non-homentropic disk and show that vortices can be created by the baroclinic instability in a 2D disk. Then we check the stability of the 2D vortices with respect to the 3D elliptic instability. Finally, we study the formation and growth of 3D vortices in a 3D stratified disk.

The paper is organized in the following way: the basic equations, the disk model and the numerical method are presented in section 2. Section 3 is devoted to the two dimensional simulations. The effect of the nature and strength of the thermal transfer on the baroclinic instability is analyzed in detail.  Three dimensional simulations are performed in section 4. The structure of the vertical stratification is first examined. Then,  
the formation of long-lived vortices in a fully 3D simulation with an unstable mid-plane layer is demonstrated. 
The results are briefly summarized and discussed  in section 5.

\section{Fluid equations and numerical method}
\label{equation}

\subsection{Standard disk assumptions}

The gas of the nebula is assumed to be a mixture of molecular hydrogen and helium with a mean molecular weight $\mu = 2.34 $g/mol. The low pressure conditions inside the disk justify the use of the perfect gas law as the appropriate equation of state.

In the steady state, the gas is stratified with an hydrostatic equilibrium in the vertical direction $z$ and a pressure modified centrifugal equilibrium in the radial direction $r$. The gas flows around the star at nearly the Keplerian velocity and fills a flared disk in which the pressure scale height is  $H(r) = c_s(r)/\Omega_K(r)$ where $c_s=\sqrt{\gamma P/\rho}$ is the sound speed and $\Omega_K=1/r^{3/2}$ is the kepler angular velocity. 

The surface density and the temperature of the disk are assumed to decrease as simple power laws of the distance to the star:  $r^{-p}$ and $r^{-q}$, respectively; numerical values follow the model of the minimum mass solar nebula (e.g. Hayashi, 1981). The self-gravity of the gas is neglected and we focus on optically thick regions of the nebula where the coupling of the gas with the magnetic field is negligible due to weak ionization. In these regions the presence of dust particles embedded in the gas also justifies the presence of thermal transfer which is a crucial ingredient for the baroclinic amplification of the vortices. The problem is addressed by solving numerically the full set of the compressible hydrodynamical equations in cylindrical coordinates with an energy equation that includes thermal transfer. 
\subsection{The governing equations}
\label{sec:equations}

The standard equations for an inviscid gas flowing around a central gravitational potential read
\begin{equation}
{ \frac{\partial \rho}{\partial t} } + {\ve{\nabla} } \cdot (\rho {\ve V}) =0
 \label{gov-eq1}
 \end{equation}
\begin{equation}
{ \frac{\partial \rho {\ve V }}{\partial t} } +{\ve\nabla} \cdot (\rho {\ve V}{\ve V}) + {\ve \nabla} P =   {\rho {\ve \nabla} \phi} 
 \label{gov-eq2}
 \end{equation}
\begin{equation}
{ \frac{\partial \rho e}{\partial t} } +{\ve \nabla} \cdot ({\ve V}(\rho e +P)) = \rho {\ve V} \cdot {\ve \nabla} \phi + Q
 \label{gov-eq3}
 \end{equation}
\begin{equation}
\rho e = {\frac P{\gamma -1}} + {\frac 1{2}}\rho {\ve V}^{2}
 \label{gov-eq4}
 \end{equation}
where  $\rho$ and $P$  are the density and pressure of the gas; $\ve V$ is the gas velocity and $ e$ its total specific energy; $\phi$ is the gravitational potential, $Q$ the heat transfer and  $\gamma = 1.4$ the adiabatic index.

The equations will be used in non-dimensional form thanks to the following normalization
\begin{subequations}
  \begin{align}
    \ve{r} & =r_o \tilde{\ve{r}}  ~~~~~~~~~~  \ve{V}=v_o \ve{\tilde{V}}  \\
   \rho & =\rho_o \tilde{\rho} ~~~~~~~  P  =\rho_o v_o^2 \tilde{P} ~~~~~~~  T= {\frac{v_o^2}R} \tilde{T}
  \end{align}
\end{subequations}
where the $_o$ index refers to the values at the reference radius $r_o$ and $R$ is the ideal gas constant; $v_o=({GM}/r_o)^{1/2}$ and $\rho_o=\rho (r_o,z=0)$ stand for the values of the velocity and the density in the mid-plane and at the radius $r_o$.\\
In the following, the tilde over the new dimensionless variables is dropped. The form of the equations  \eqref{gov-eq1}-\eqref{gov-eq4} is unchanged, except the gravitational potential which now reads $\phi=1/(r^2+z^2)^{1/2}$. The important parameter is the Mach number $M_{A_o} = v_o/\sqrt{\gamma R T_o}$ where $T_o$ is the gas temperature at radius $r_o$.

\subsection{Equilibrium state of the disk}

In 2D the steady state is assumed to be simple power law in temperature  and  surface density\\
\begin{equation}
T_D=  \frac{r^{-q}}{\gamma M_{A_o}^2} ~~;
\label{equiT}
\end{equation}
\begin{equation}
\Sigma_D=  r^{-p} ~~
\end{equation}
where $p=1.5$ and $q=0.5$ in the case of the standard MMSN hypothesis.
The perfect gas equation of state gives\\
\begin{equation}
P_D=  \Sigma_D T_D=\frac{r^{-p-q}}{\gamma M_{A_o}^2} ~~.
\end{equation}
The radial velocity is assumed to be null and the azimuthal velocity is given by the radial hydrodynamic equilibrium
\begin{equation}
  v_D  = \sqrt{r{ \frac{\partial \phi}{\partial r}} + {\frac r{\Sigma_D} }{ \frac{\partial P_D}{\partial r}} } ~ .
\end{equation}
For large Mach number, the disk angular velocity $\Omega_D=v_D/r$ is close to the Kepler profile $\Omega_K$ in the mid-plane.
This equilibrium solution only depends on a small number of parameters: the indices $p$ and $q$, and the Mach number $M_{A_o}$. It is also interesting to note that the structure of the disk is independent of the disk mass.\\
The pressure scale height is given by
\begin{equation}
H(r) =\frac{c_s}{\Omega_K}=\frac{ r^{(3-q)/2}}{ M_{A_o}}.
\label{exp:HD}
\end{equation}

\subsection{Radial stratification}
\label{sec:initial} 

In the radial direction, disk stratification is measured by the Brunt-V\"ais\"al\"a frequency
\begin{equation}
N_r^2=-{ \frac1{\gamma\rho}} { \frac{\partial P}{\partial r}}{ \frac\partial {\partial r}} \ln\left ( \frac P{\rho^\gamma}\right) ~,
\label{Nr}
\end{equation}
and the Schwarzschild criterion states that this stratification is stable if $N_r^2>0$ and unstable if $N_r^2<0$.

In a two dimensional disk at equilibrium, pressure and density can be expressed as simple power-laws and the Brunt-V\"ais\"al\"a frequency writes
\begin{equation}
N_r^2= \frac{(p+q)(p(\gamma-1)-q)}{\gamma r^2}T  ~~.
\end{equation}
Then, if pressure and density are decreasing outward (i.e $p$ and $q$ are positive), we find that stratification is unstable when $p<q/(\gamma-1)$. In term of the Richardson number $Ri={N_r^2 / S^2}$ , where $S= -3\Omega_K /2$ is the Kepler shear, the condition of instability is then equivalent to
\begin{equation}
Ri>0  ~~.
\end{equation}

\subsection{Thermal transfer}
\label{sec:transf} 

Thermal transfer in optically thick regions of the nebula results from an energy dissipation either in the form of a black body radiation escaping from the disk or a diffusive process through the scattering of IR photons in the disk. These complex phenomena are not considered here. In the present paper, we mainly  focus on the physical consequences of thermal transfer. As commonly done, we assume that thermal transfer only takes place in the form of thermal relaxation or heat diffusion. Moreover, we  clearly distinguish between the two cases by assuming either pure relaxation or pure diffusion.\\

$\bullet$ Thermal relaxation\\
In this case the gas temperature variations are assumed to relax toward an equilibrium temperature $T_0$ at a rate
 \begin{equation}
 Q =  {\frac\gamma{\gamma-1}} {\rho}   { \frac{(T-T_o)}\tau }
 \label{relax}
 \end{equation}
where $\tau$ is the cooling time normalized to the inverse of the orbital frequency. Thermal relaxation then affects equally  all spatial scales. However, it does not conserve the total energy which is now a function of time.\\

$\bullet$ Heat diffusion\\
The thermal energy of the gas is assumed to diffuse in the disk at a rate
 \begin{equation} 
 Q =  {\frac\gamma{\gamma-1}}  {\frac{ \rho}{Pe}}   {\Delta (T-T_o) }   
 \label{diffu}
 \end{equation}
 where $\Delta$ is the Laplacian, and $Pe$ is the P\'eclet number, a dimensionless number that quantifies the importance of the transfer in the disk and  defined as
\begin{equation} 
Pe =  { \frac{{H(r_*)~{r_*} \Omega(r_*)}}\kappa}. 
\label{peclet}
\end{equation}
where $\kappa $ is a diffusion coefficient and $r_*$ the mean orbital radius of the computational box (in what follows we will use $Pe/ \rho(r_*)$ instead of $Pe$).  Heat diffusion operates at all  spatial scales  but its effect is stronger on smaller scales.  We shall see the consequences of this property on the evolution and structure of large size vortices. We also note that, in contrast to thermal relaxation, heat diffusion conserves total energy.

\subsection{Numerical method}
The system of non-linear equations (1)-(4) is solved using a second order finite volume scheme, the MUSCL Hancock scheme, and an exact Riemann solver. The two dimensional version of the code is described in \citet{Inaba2005} (see also C. Surville, 2013). The three dimensional version is the same as used in \citet{Richard2013}. The two versions were used in this paper to study the baroclinic instability in a two-dimensional or a three-dimensional context. We do not use the standard shearing sheet approximations and our simulations correspond to local integrations of the full set of the fluid equations.

No explicit dissipation mechanism is present in our simulations. Viscous effects (numerical diffusion) originate either at a global scale owing to the numerical 
interpolation of the field on a finite size grid, or at small scales, owing to the existence of a cut-off length in numerical integration. They can be described with effective viscosities: $\nu_{a}$ at  large scales and $\nu_{n}$ at  small scales. The first one has been estimated by analyzing the time evolution of known viscous solutions \eg \citet{Pringle81}; 
numerical benchmarks with the 2D version of our code (Surville, PhD thesis) have shown that $\Delta \Sigma/(\Delta t \Sigma\Omega)\simeq 3.2~10^{-6}$ so that $\nu_a \simeq 7.2~10^{-6} r_*^2\Omega_* \simeq 10^9 m^2 s^{-1}$ (or $Re\simeq 5.2~10^5$).
The second one scales as $\Omega_* \delta_*^2 $ where $\delta_*$ is the mesh size; in the radial direction, this estimate gives $\nu_{n}\simeq 10^{10} m^2s^{-1}$  (or $Re\simeq5.2~10^4$).

\section{Two dimensional simulations of the baroclinic instability}
\label{2D}
In this section, we perform  two-dimensional simulations to recover and extend a few results already obtained in the literature. 
As explained in LP10 the necessary conditions for the subcritical baroclinic instability (SBI) are: (i) a radially unstable entropy gradient or negative Richardson number, (ii) a thermal transfer at a rate that can sustain the baroclinic generation of vorticity, (iii) a seeding of the disk with finite amplitude perturbations to trigger the non-linear response of the disk.

To satisfy the above requirements our 2D simulations are performed with the following initial conditions: \\
(a) a background temperature with an index $q=0.5$; \\
(b) a fixed value of the Mach number $M_{A_o}=8$ (at $r_o$) or correspondingly a scale height $H=1.55$ at $r=7.5$ ; \\
(c) a perturbation of the equilibrium state obtained by randomly superimposing on the background density small gaussian bumps with amplitude $\sim10\%$ .

The computational domain of the simulations was limited in the radial direction to $7<r<8$ and in the azimuthal direction to $0<\theta<\pi/4$. The numerical resolution was set up to 300$\times$300, that is  465 cells by scale height in the radial direction and 80 cells by scale height in the azimuthal direction. The results of the simulations are presented below. 

\subsection{A two stage development}

\begin{figure*}
\begin{center}
\includegraphics[width=0.5cm,height=2.9cm]{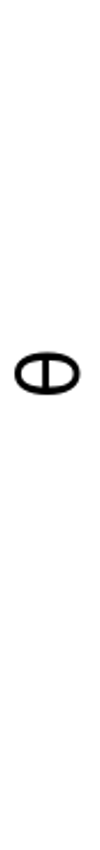}
\includegraphics[width=3.9cm]{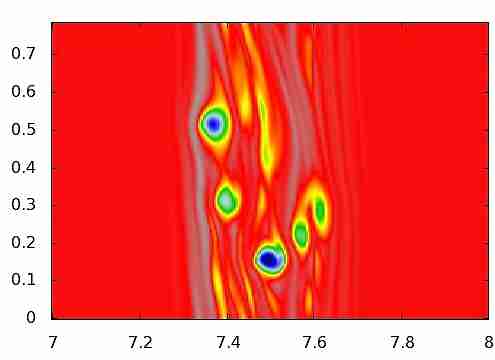}
\includegraphics[width=3.9cm]{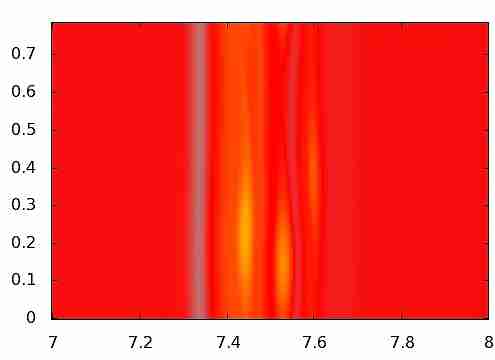}
\includegraphics[width=3.9cm]{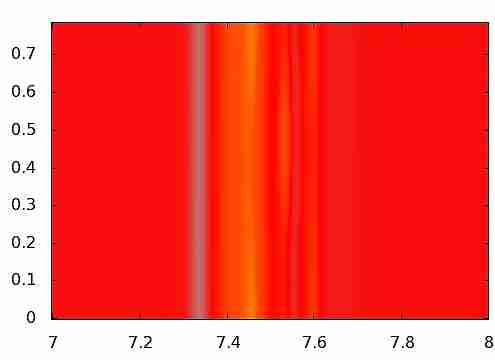}
\includegraphics[width=3.9cm]{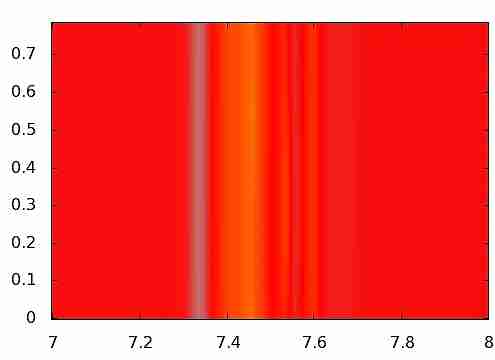}
\includegraphics[height=2.9cm]{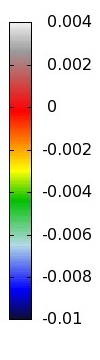}\\
\includegraphics[width=0.5cm,height=2.9cm]{theta.png}
\includegraphics[width=3.9cm]{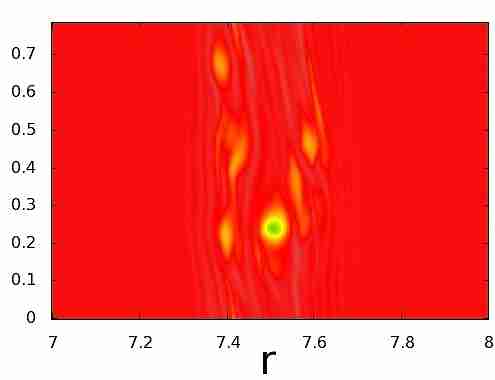}
\includegraphics[width=3.9cm]{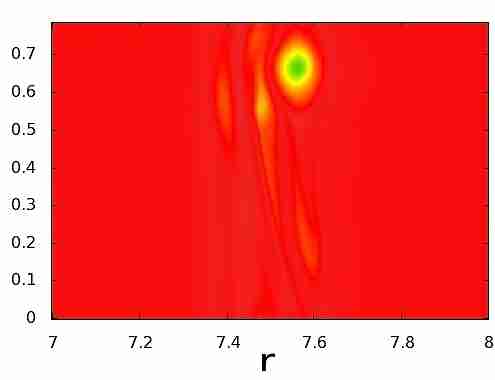}
\includegraphics[width=3.9cm]{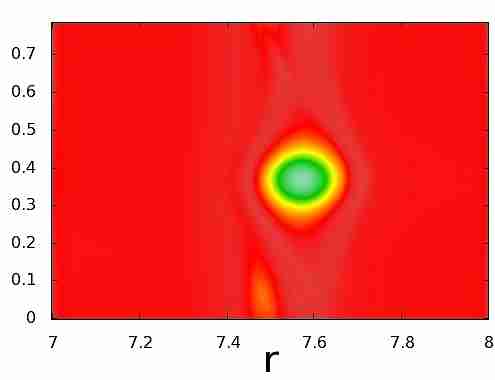}
\includegraphics[width=3.9cm]{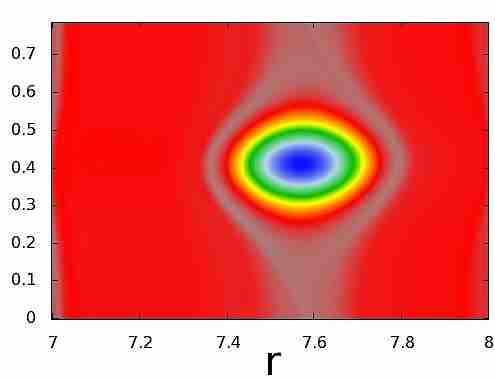}
\includegraphics[height=2.9cm]{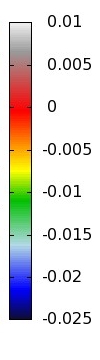}
\end{center}
 \caption{Vorticity field after 20, 200, 400 and 600 rotations (from left to right, respectively) in a 2D disk with thermal relaxation with a cooling time $\tau=160$. Top row: stable stratification ($p=1.5$, $q=0.5$). Bottom row: unstable stratification ($p=1$, $q=0.5$). Vortices are rapidly  decaying in the case of a stable stratification  while they are slowly but systematically amplified in the case of an unstable stratification.}
\label{Evolution}       
\end{figure*}

In order to illustrate the development  of the baroclinic instability in two-dimensional disks, we have first carried out simulations with thermal relaxation. The crucial importance of radial stratification was tested by comparing simulations performed for background densities with significantly different radial gradients. We have chosen two power-laws with different values of the index $p$.

For $p=1.5$, stratification is stable since the Richardson number is positive ($Ri\approx 2.10^{-3}$ at $r=7.5$) whereas for $p=1$, stratification is unstable since the Richardson number is negative ($Ri\approx -1.5.10^{-3}$ at $r=7.5$) . 
Baroclinic feedback is expected only in the case of unstable stratification that is when $p < q/(\gamma-1)$ (here $p<1.25$).

In Fig. \ref{Evolution}, the vorticity  is plotted  at different instants of the simulation. Two different stages can be distinguished in the evolution of the vorticity: a first one during which small vortices are formed from the initial density perturbations, nearly irrespective of stratification; a second one, after approximatively  50 rotations, during which the evolution strongly depends on stratification. If stratification is stable, the small vortices formed during the first stage gradually decay in a few hundred rotations whereas, if stratification is unstable, they are amplified and grow as a function of time. These observation confirms known results.\\

\begin{figure}
\begin{center}
\includegraphics[width=9cm]{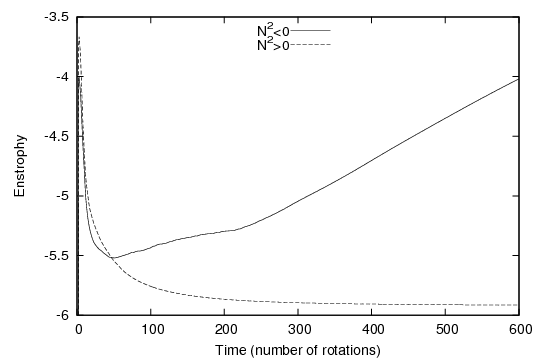}
\end{center}
 \caption{Evolution of the enstrophy in a 2D stratified disk with thermal relaxation. The full line corresponds to an unstable stratification  ($N_r^2<0$) whereas the dotted line corresponds to a stable stratification ($N_r^2>0$). Enstrophy is computed for $\tau=160$.}
\label{enstr-strat}       
\end{figure}

Figure \ref{enstr-strat} shows the variation of the enstrophy $\iint \omega_z^2 dS /2$ as a function of time. This enstrophy evolution  is consistent with the evolution of the vorticity field. The peak at the very beginning corresponds to the formation of vortices from the initial density seeds. The rest of the plot corresponds either to a gradual decay of the vortices or to their baroclinic amplification depending on whether stratification is stable or not, respectively.

The baroclinic amplification of the vortices (thereafter BVA) was first identified by Petersen et al. (2007b).  LP10 also explained  why unstable disk stratification is required for the vortex amplification.  
The mechanism operates along the cyclic path of a gas parcel around the vortex center; it consists in a thermalization of the parcel during its azimuthal excursion followed by an acceleration of the parcel during its radial excursion. 
It must be noted that, in the case of a stable stratification, the mechanism becomes ineffective since the radial motion of the parcel is not accelerated but, on the contrary, decelerated; this leads to a gradual weakening of the vortex instead of an amplification. This explains why vortices weaken when $N_r^2>0$. We want to stress that this gradual decay of the vortices is very different from the breakdown of the vortex which may occur in three-dimensional disks due to the elliptic instability.
In the rest of the section, we only consider an unstable stratification by fixing the parameter $p$ and $q$ to the values $p=1.5$ and $q=0.5$. Our goal is to analyze the importance of the thermal transfer in the baroclinic instability. We consider first thermal relaxation then heat diffusion. 

\subsection{Thermal relaxation} 
\begin{figure}
\begin{center}
\includegraphics[width=9cm]{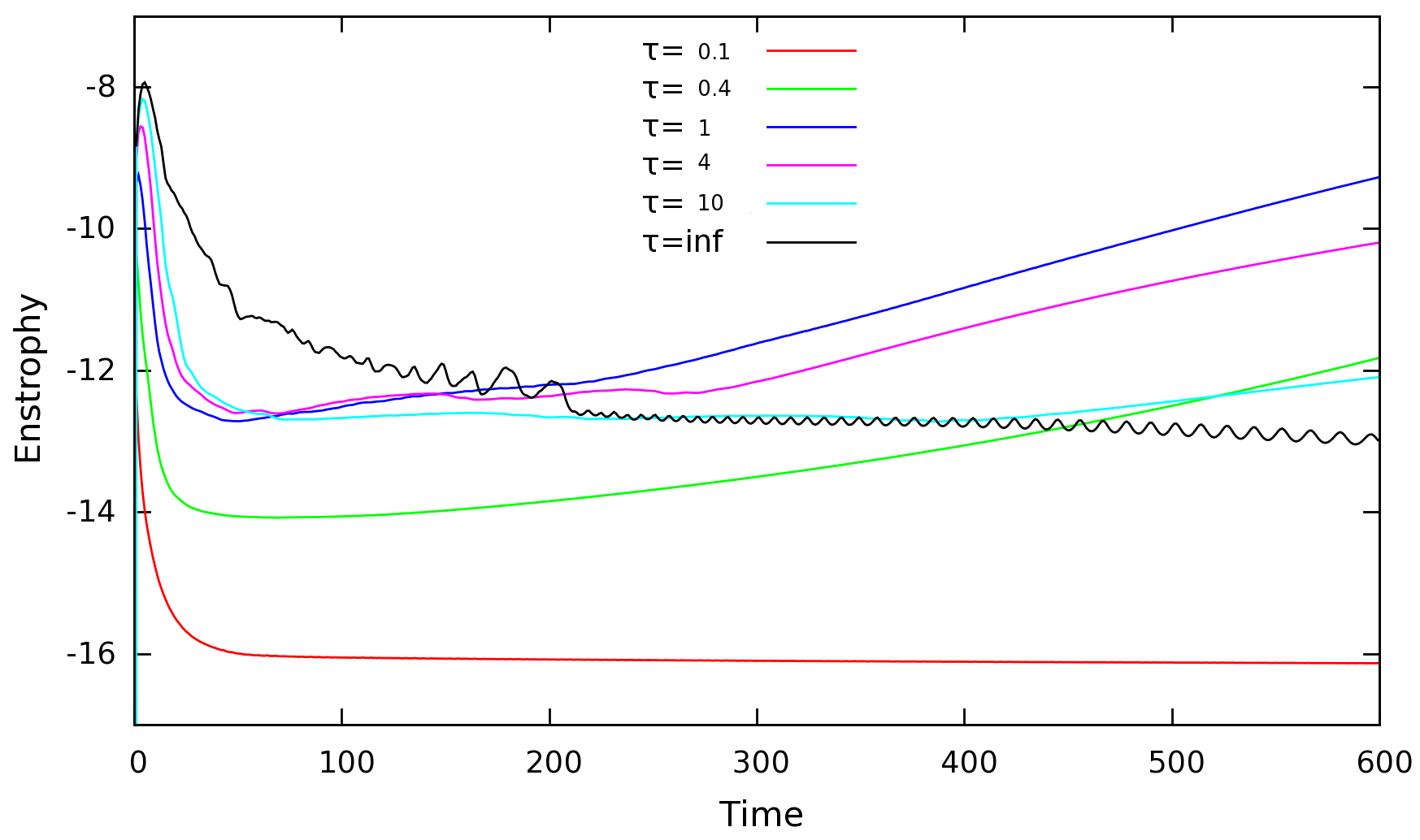}
\end{center}
 \caption{Enstrophy as a function of time, in a 2D stratified disk ($p=1.5$, $q=0.5$) with thermal relaxation for different values of the cooling time. Only intermediate values of $\tau$ lead to an efficient vortex amplification. }
\label{enstrophie}       
\end{figure}

The results are presented in Fig. \ref{enstrophie} which shows the evolution of the enstrophy as a function of time and for different values of the cooling time $\tau$. The cooling time is given in term of the orbital period at $r=7.5$. In the beginning of the simulation, the production of vorticity is very effective with the formation of a peak of enstrophy which is associated with the birth of small vortices from the density perturbations initially seeded in the disk. We also observe that  the longer the cooling time, the higher the initial peak of enstrophy. The generation of vortices is therefore more efficient for longer cooling times.\

The subsequent evolution of the enstrophy depends on the cooling time. For both short and large cooling times, enstrophy decreases monotonically while for intermediate values it increases after the rapid phase of decay. 
 For intermediate cooling time, the growth of enstrophy corresponds to  a baroclinic amplification of the vortices.  This amplification is effective on a time scale much longer (some hundred rotations) than the time necessary 
 for the creation of the vortices. 
During this amplification phase, vortices are gradually merging one another to give a single vortex that continues to grow until it reaches the maximum size compatible with the limits of the computational box.\\

\begin{figure}
\begin{center}
\includegraphics[width=9cm]{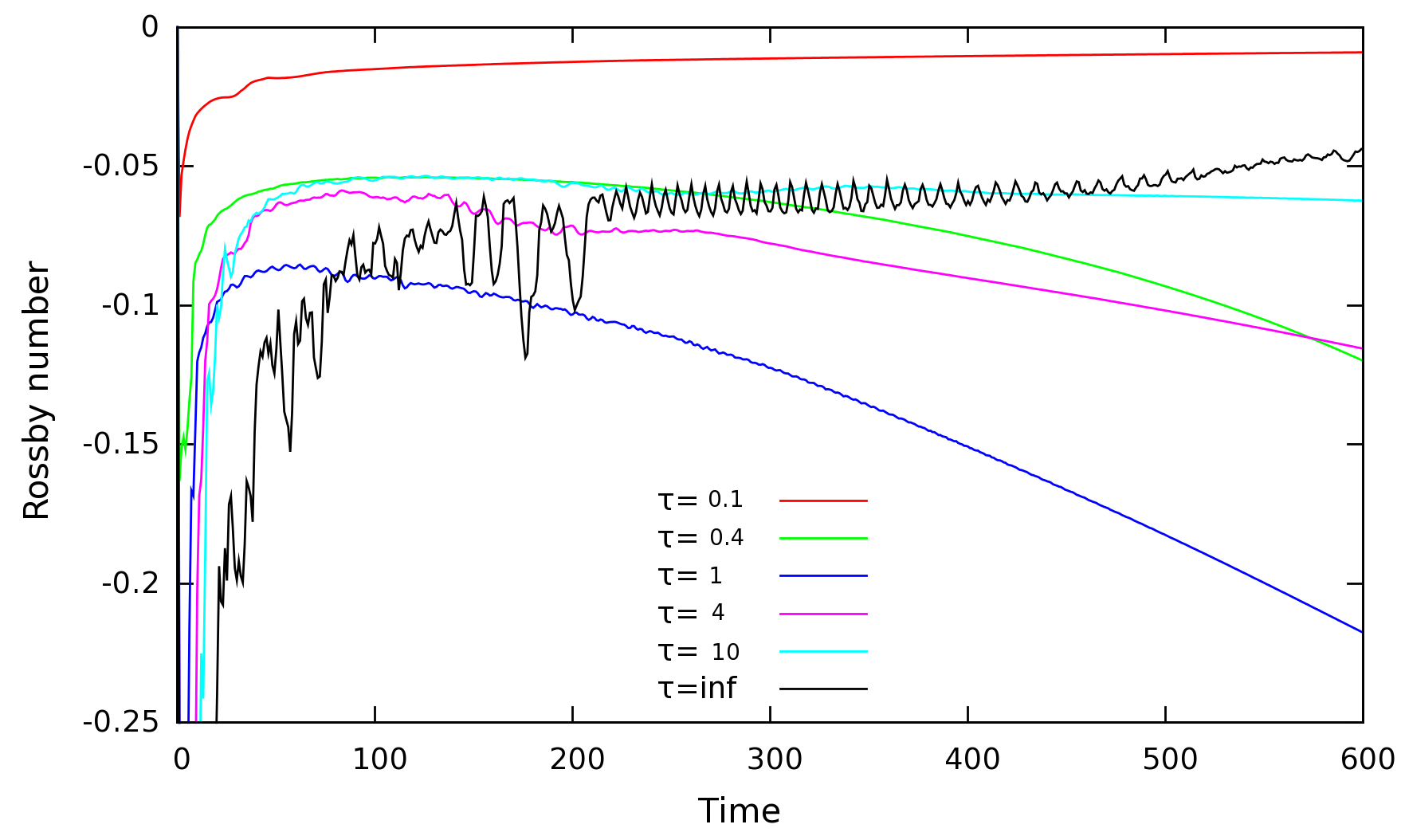}
\end{center}
 \caption{Rossby number of the strongest vortex as a function of time  in a 2D stratified disk ($p=1.5$, $q=0.5$) with thermal relaxation for different values of the cooling time.}
\label{Ro-tau}       
\end{figure}

\begin{figure}
\begin{center}
\includegraphics[width=9cm]{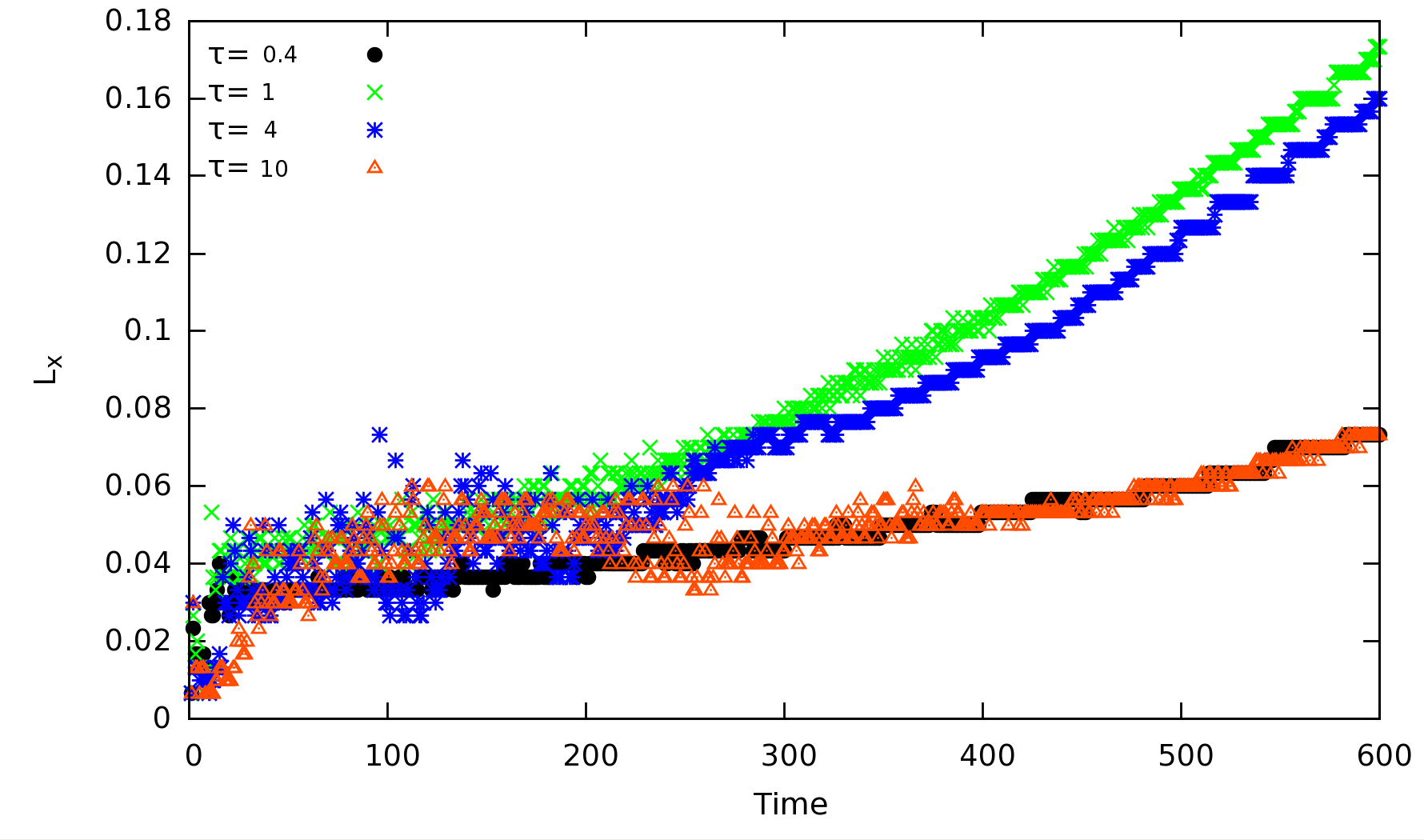}
\end{center}
 \caption{Radial extent of the strongest vortex as a function of time in a 2D stratified disk ($p=1.5$, $q=0.5$) with thermal relaxation for different values of the cooling time.}
\label{lx}       
\end{figure}

\begin{figure}
\begin{center}
\includegraphics[width=9cm]{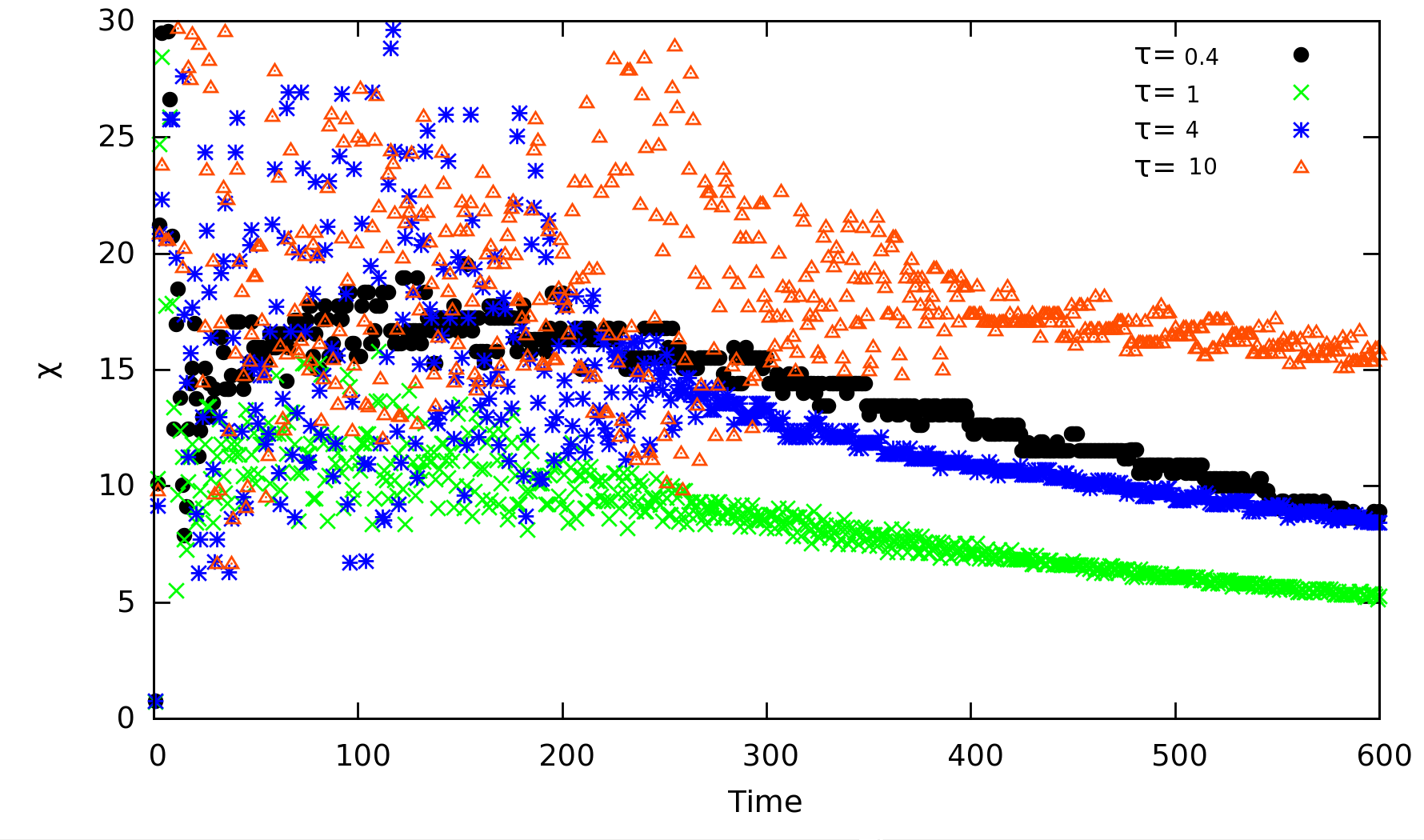}
\end{center}
 \caption{Aspect ratio of the strongest vortex   as a function of time in a 2D stratified disk ($p=1.5$, $q=0.5$) with thermal relaxation for different values of the cooling time.}
 \label{chi}       
\end{figure}

\begin{figure}
\begin{center}
\includegraphics[width=9cm]{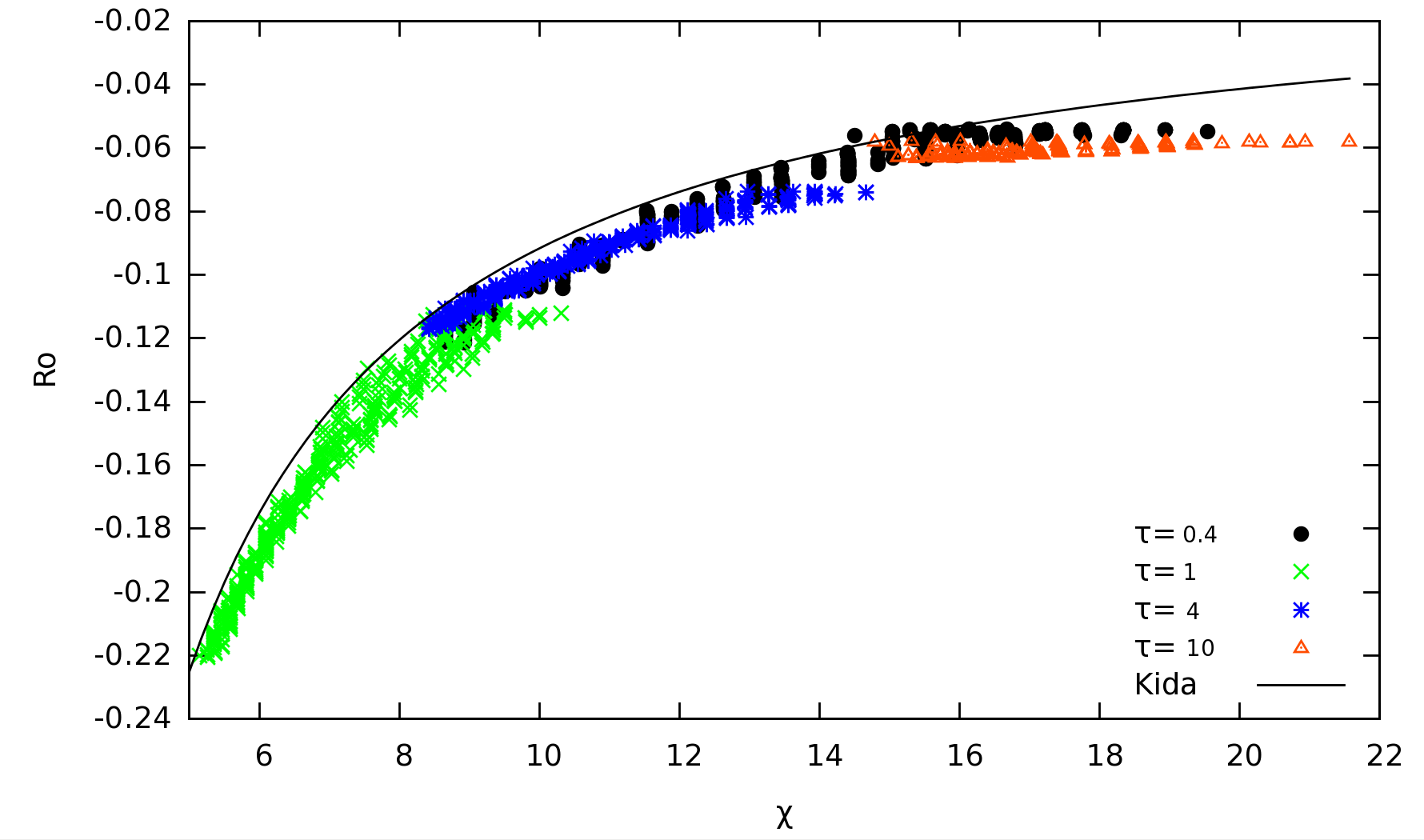}
\end{center}
 \caption{Rossby number of the strongest vortex as a function of its aspect ratio   in a 2D stratified disk ($p=1.5$, $q=0.5$) with thermal relaxation for different values of the cooling time and different instants.  
 The black curve represents the relation  (\ref{Kida}) obtained for Kida's vortex model.}
\label{chi-Ro}       
\end{figure}

When $\tau=\infty$ or $\tau =0.1$ no baroclinic amplification is observed. In the first case the flow is adiabatic (no thermal transfer) and we find that the small vortices formed during the first phase cannot grow. This confirms that thermal transfer is required to get a baroclinic amplification. It is, however, interesting to stress that the adiabatic assumption does not inhibit vortex formation but only the amplification mechanism.
In the second case ($\tau=0.1$) enstrophy does not increase because no vortex has emerged during the first phase. No baroclinic amplification is possible since no vortex is present. 
To summarize, the formation of large and strong vortices is  possible for intermediate values of $\tau$ ($\tau \approx \Omega_K^{-1}$) only.\
After studying the global production of vorticity with the enstrophy, it is also interesting to look at the growth of a single vortex. For this purpose we follow the evolution of the strongest vortex formed at the end of the birth period. The strength of this vortex is estimated using the Rossby number\\
\begin{equation}
Ro= {\frac{\omega_z}{2 \Omega_D(r_v)}}
\end{equation}
where $\omega_z$ is maximum relative vorticity of the vortex and $r_v$ is the radial position of the vortex.
Figure \ref{Ro-tau} shows the evolution of the Rossby number as a function of time and for various cooling times.
We clearly see the strength increase of the vortex  when the baroclinic amplification is active. The strongest vortex also increases in size. This is illustrated in Fig. \ref{lx} where is shown the evolution of the radial extent 
of the strongest vortex during its amplification.\
The aspect ratio $\chi$ of the vortices decreases as a function of time (see Fig. \ref{chi}), as expected from the systematic increase of the vorticity. The strong fluctuations observed in the beginning of the evolution are due to the difficulty to evaluate the aspect ratio of the vortices as they are merging with one another. After some hundred rotations, a single vortex remains in the disk and the dispersion of the measures strongly reduces, uncovering a slow decrease of $\chi$ as a function of time.\\
Figure \ref{chi-Ro} shows the Rossby number as a function of the aspect ratio and for various cooling times. A clear relation exists between the two parameters, nearly independent of the cooling time. It is well fitted by the analytical expression derived for Kida's vortex model \citep{Kida81}
\begin{equation} \label{Kida}
Ro =-{\frac 3 {4}} {\frac{\chi +1}{\chi(\chi-1)}}.
\end{equation}
\subsection{Heat diffusion}
\begin{figure}
\begin{center}
\includegraphics[width=9cm]{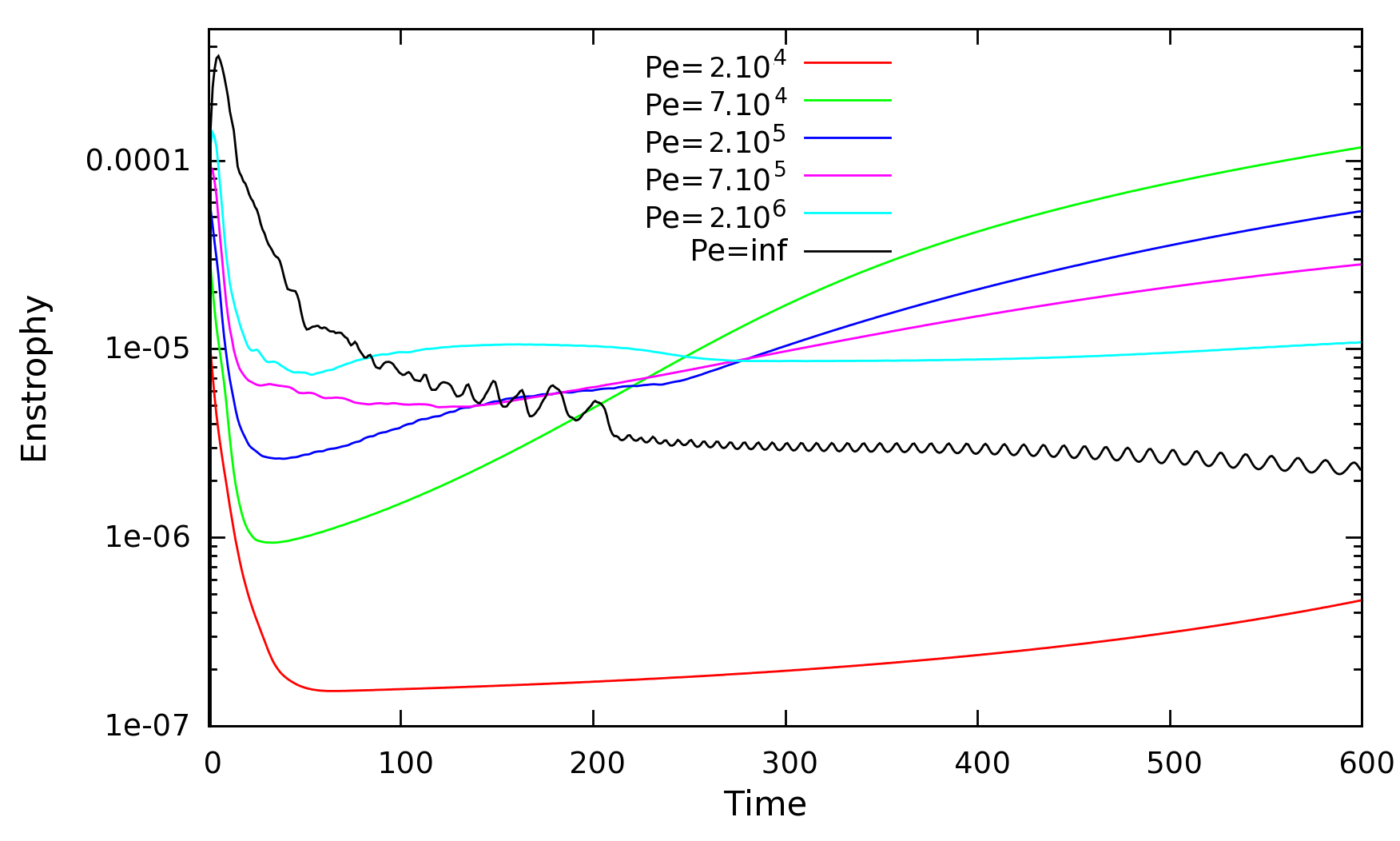}
\end{center}
 \caption{Enstrophy as a function of time in a 2D stratified disk, in the case of heat diffusion and for different values of the P\'eclet number.}
\label{Enstro_Pe2}       
\end{figure}

In this subsection, thermal transfer is due to heat diffusion. The strength  of heat diffusion is quantified by the P\'eclet number.
The smaller this number, the stronger the diffusion. The non-diffusive case corresponds to an infinite P\'eclet number. 
The  computational domain has been chosen to be a box with $7<r<8$ and $0<\theta<\pi/2$ except in the simulation presented in Fig. 9 in which the box is extended to $6.5<r<8.5$ and $0<\theta<\pi$ . 
The results we get for heat diffusion are quite similar to those obtained  for thermal relaxation. Two stages can also be distinguished in the evolution of the enstrophy plotted in Fig. \ref{Enstro_Pe2}. The peak at the very beginning corresponds to a first stage with the formation of small vortices from the density bumps initially seeded in the disk. The subsequent evolution corresponds to a second stage during which vortices are growing under the baroclinic feedback mechanism. Our simulations show that baroclinic amplification is effective only for large enough values of the P\'eclet number (Pe $> 2.10^4$). However, as for thermal relaxation, the amplification is weakened for very large $Pe$ and disappears at $Pe=\infty$. There is therefore an optimal heat diffusion for the amplification which corresponds to $Pe \approx 10^5$.\

\begin{figure}
\begin{minipage}{.92\linewidth}
\centering
\includegraphics[width=0.4cm,height=3cm]{theta.png}
\includegraphics[width=3.8cm]{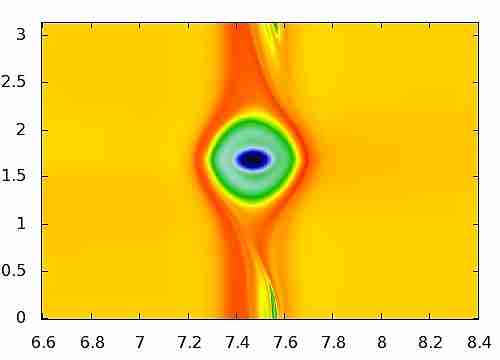}
\includegraphics[width=3.8cm]{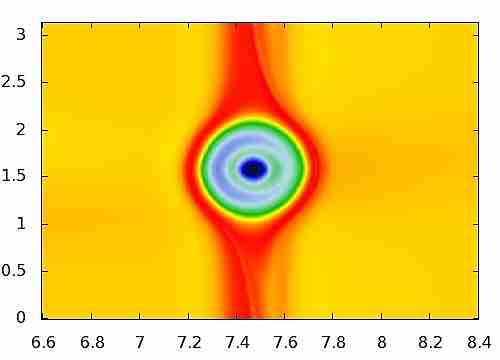}
\includegraphics[width=0.4cm,height=3cm]{theta.png}
\includegraphics[width=3.8cm]{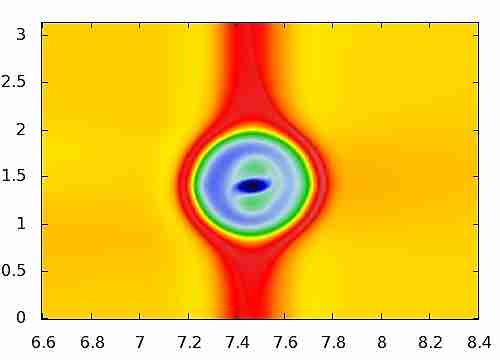}
\includegraphics[width=3.8cm]{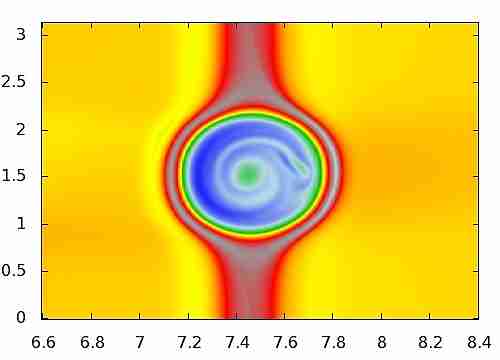}
\includegraphics[width=0.4cm,height=3cm]{theta.png}
\includegraphics[width=3.8cm]{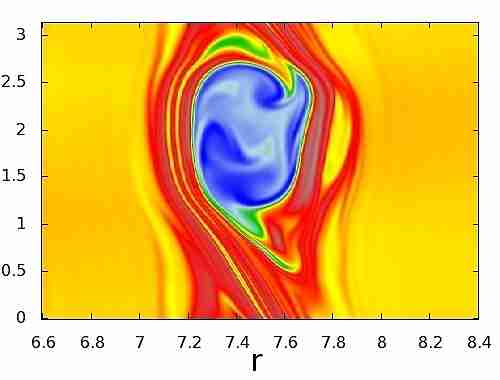}
\includegraphics[width=3.8cm]{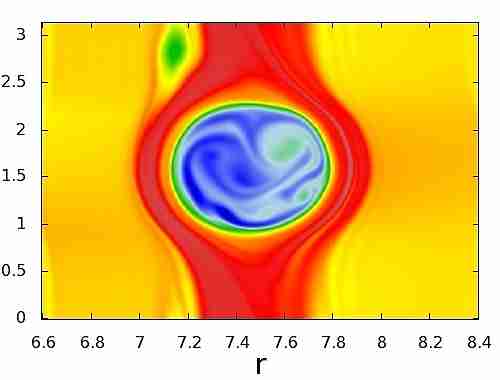}
\end{minipage}
\begin{minipage}{0.06\linewidth}
\includegraphics[width=1cm,height=6cm]{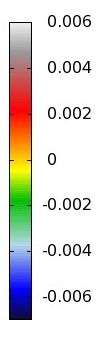}
\end{minipage}
\caption{Vorticity map in a 2D stratified disk, in the case of heat diffusion after 300, 400, 500, 700, 860 et 960 rotations, from left to right and from top to bottom. A hollow vortex forms and evolves in a vortex with a turbulent core. }
\label{diffusion2}       
\end{figure}

The main differences between an evolution under thermal relaxation and under heat diffusion clearly appear during the second stage, after the vortices have grown and merged into a single large vortex. Vorticity during this time period is presented in Fig. \ref{diffusion2} in the form of 6 different panels corresponding to 300, 400, 500, 700, 860 and 960 rotations, respectively. In the three first panels a strong peak develops in the center, surrounded by an annular bump. It must be noted that this peculiar structure looks like the one observed by \cite{Raettig2013} in their figure 8. In the fourth panel, after 700 rotations, a hollow vortex has formed : the negative vorticity in the center becomes lower than the negative vorticity on the periphery. In the two last panels the hollow vortex is strongly perturbed in the center but not in the outer rims that preserves a coherent shape. At this stage the hollow vortex is transformed in a vortex with a turbulent core that we observed till the end of the simulations.
This is a significantly different evolution than in the case of thermal relaxation.

\begin{figure}
\centering
\includegraphics[width=0.4cm,height=2.5cm]{theta.png}
\includegraphics[width=4cm]{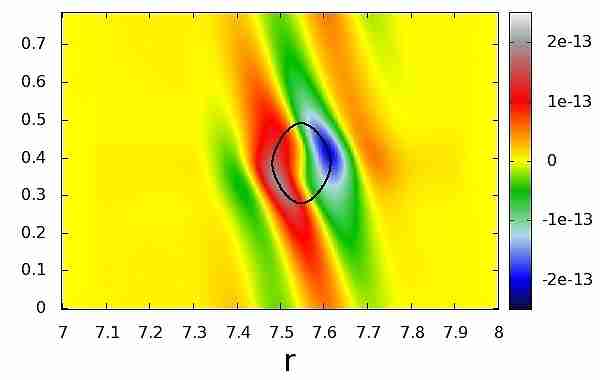}
\includegraphics[width=4cm]{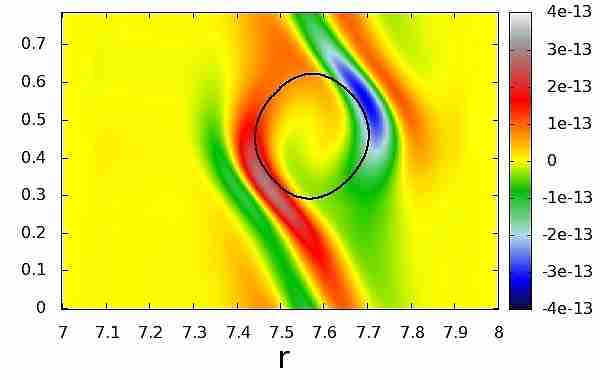}
\caption{Map of the heat transfert ($Q$) in a 2D stratified disk in the case of heat diffusion after 250 and 500 rotations, from left to right, respectively. } 
\label{Q}       
\end{figure}

The differences observed in the evolution of the vortices lie in the details of the amplification mechanism.  
Indeed, BVA can work if the gas warms up when moving azimuthally in the inner side (the side of the star) and cools down when moving azimuthally in the outer side (opposite side of the star). Inside a large size vortex, this condition is always satisfied in the case of thermal relaxation but not in the case of heat diffusion. In the presence of heat diffusion, energy exchanges are not uniform inside in the vortex. 
This point is illustrated in Fig. \ref{Q} which displays the distribution of heat transfert ($Q$) inside a growing vortex at two different instants,  after 250 and 500 orbital periods.  This figure shows that when the vortex reaches a large size, the distribution of thermal transfer changes sign in the vortex core while it remains unchanged in the vortex border. 

So, the baroclinic amplification mechanism is still active in the vortex periphery, but no more in the vortex core where an opposite mechanism now leads to the damping of the vorticity.  This could explain the formation of  hollow vortices. These hollow vortices  turn out to be unstable as observed in the last two panels of Fig. \ref{diffusion2}. 

This simulation shows that, in the case of heat diffusion, the structure of a baroclinic vortex depends on its spatial extent and evolves in a complex way. At the end of the simulations the vortex fills a large part of the available computational domain and stabilizes in a vortical structure with a big turbulent core.

\section{Three dimensional simulations}
\label{sec:baroc} 
In this section we  leave the 2D setup and consider 3D effects. 
We fixe the nature of the thermal transfer to thermal relaxation with a cooling time $\tau = 1$. 
Previous 3D simulations \citep{Lesur2010,Lyra2011} were carried out using Boussinesq approximation and constant radial stratification. 
Here we use a fully compressible code first in an vertically unstratified configuration and  then in a non-uniformly stratified configuration.  In the $z$ direction boundary conditions are set up by imposing the values derived from the equilibrium state of the disk.
\subsection{Vertically unstratified disk}
\label{sec:ellip} 
\begin{figure*}
\begin{minipage}[c]{.95\linewidth}
\includegraphics[width=0.97cm]{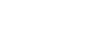}
\includegraphics[width=3.2cm]{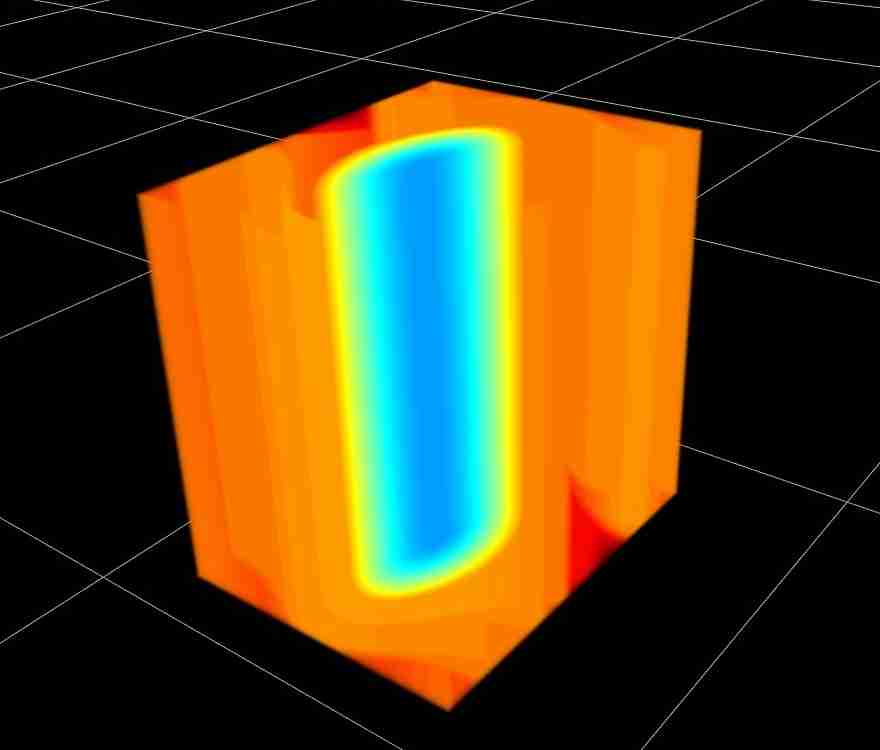}
\includegraphics[width=3.2cm]{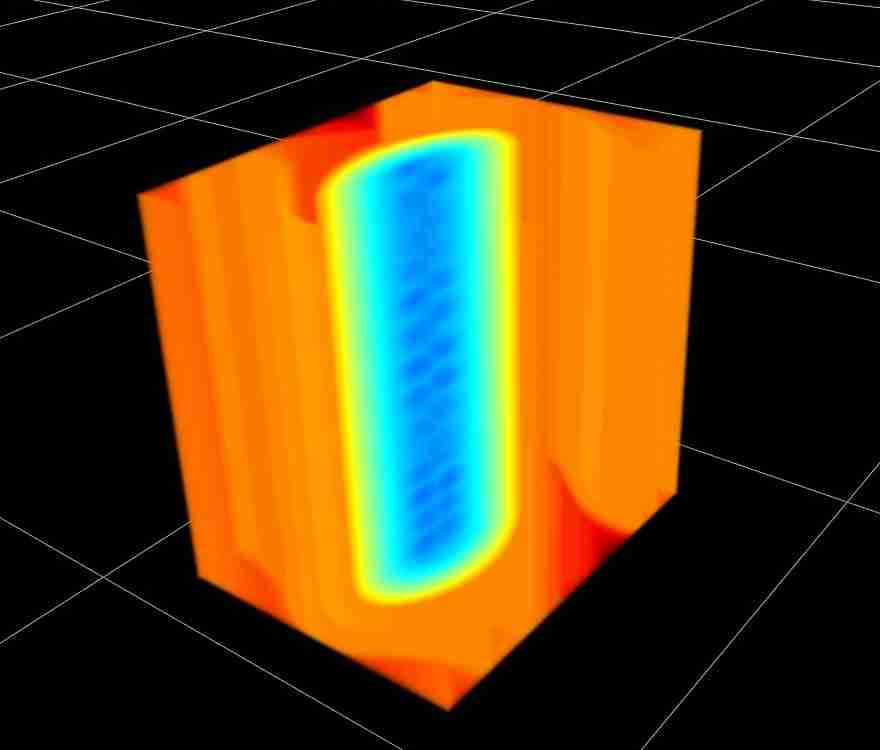}
\includegraphics[width=3.2cm]{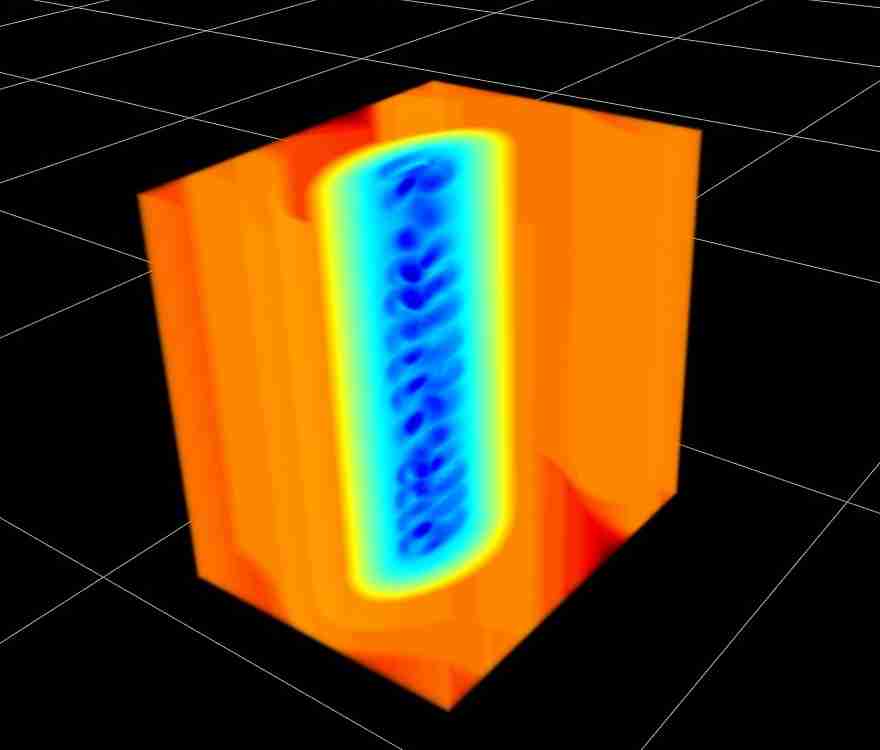}
\includegraphics[width=3.2cm]{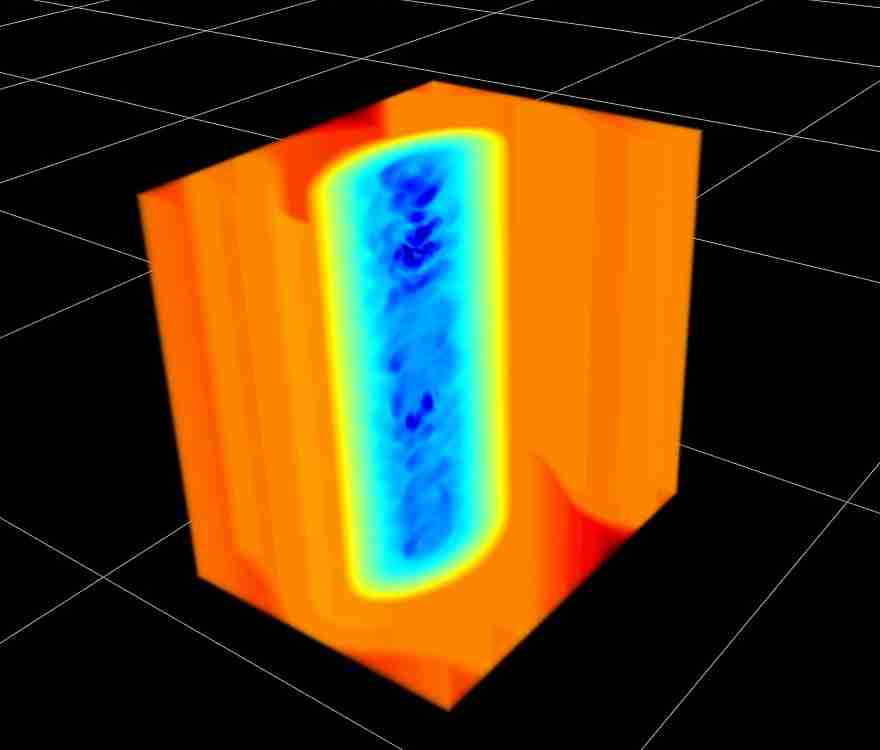}
\includegraphics[width=3.2cm]{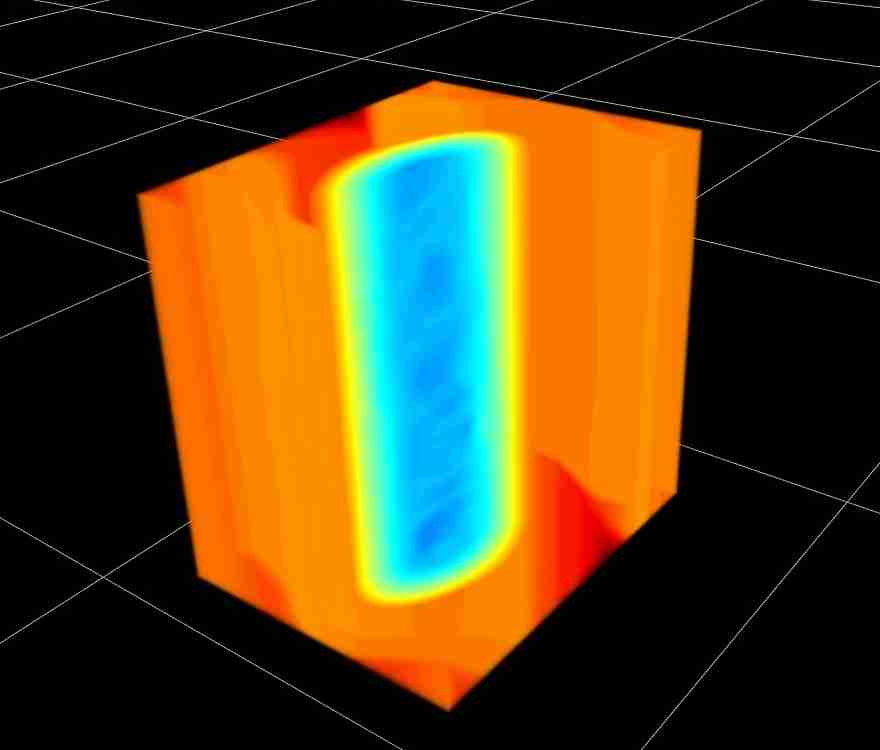}\\
\includegraphics[width=0.97cm]{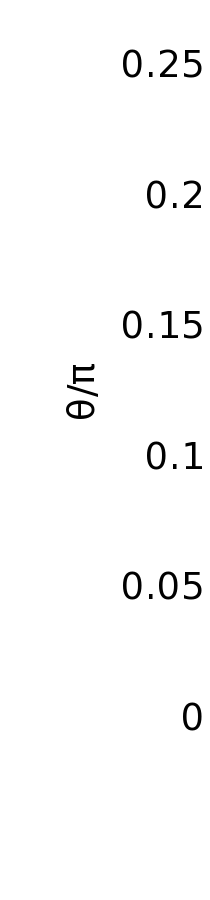}
\includegraphics[width=3.2cm]{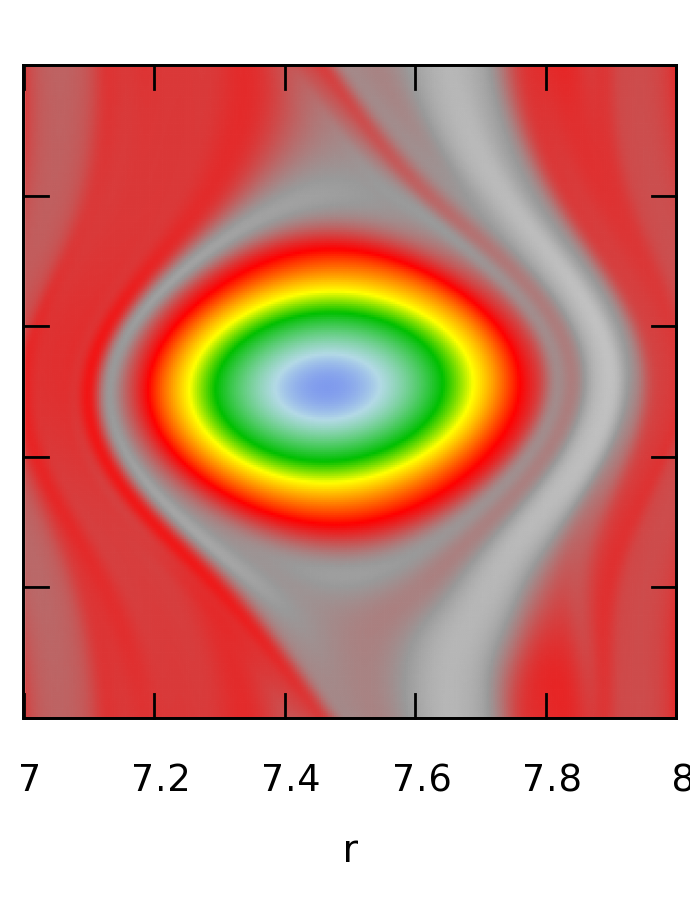}
\includegraphics[width=3.2cm]{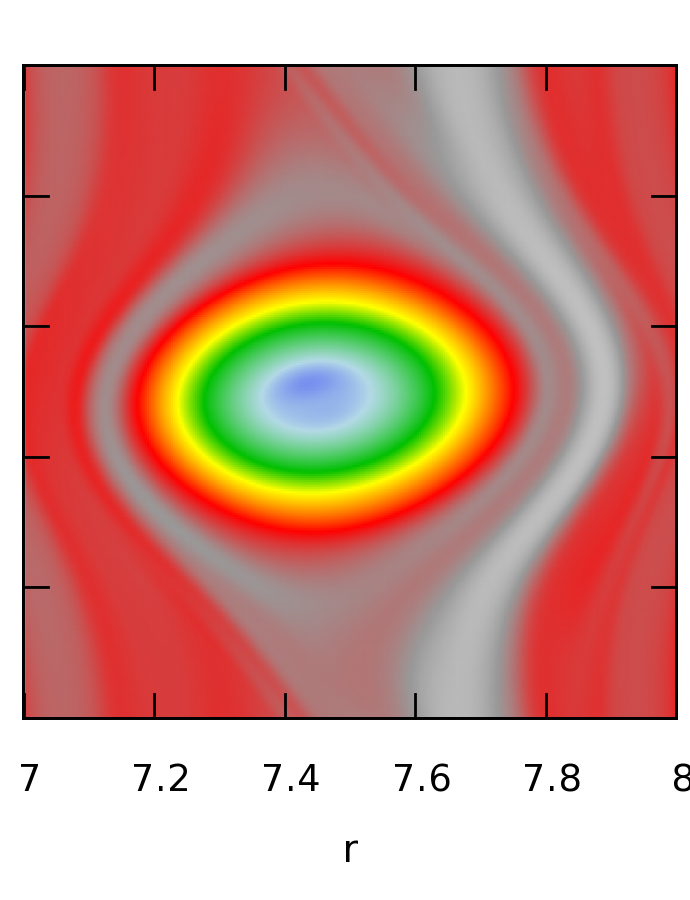}
\includegraphics[width=3.2cm]{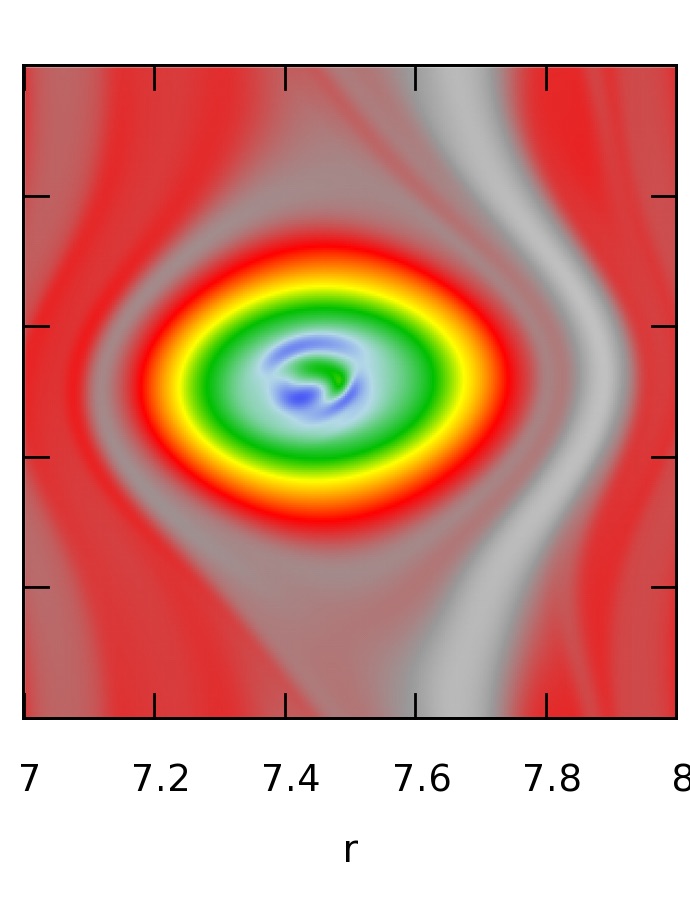}
\includegraphics[width=3.2cm]{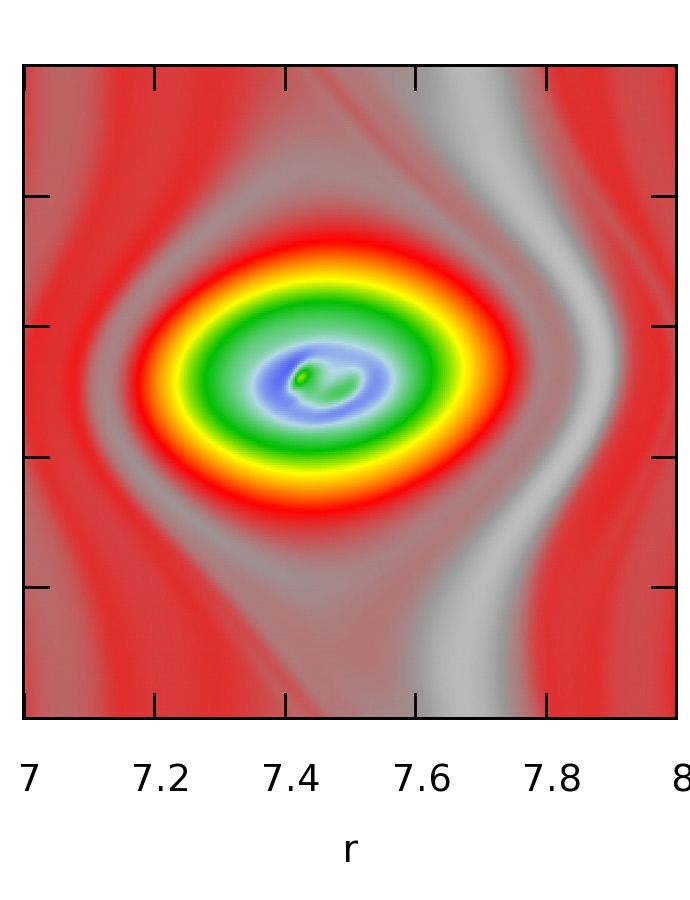}
\includegraphics[width=3.2cm]{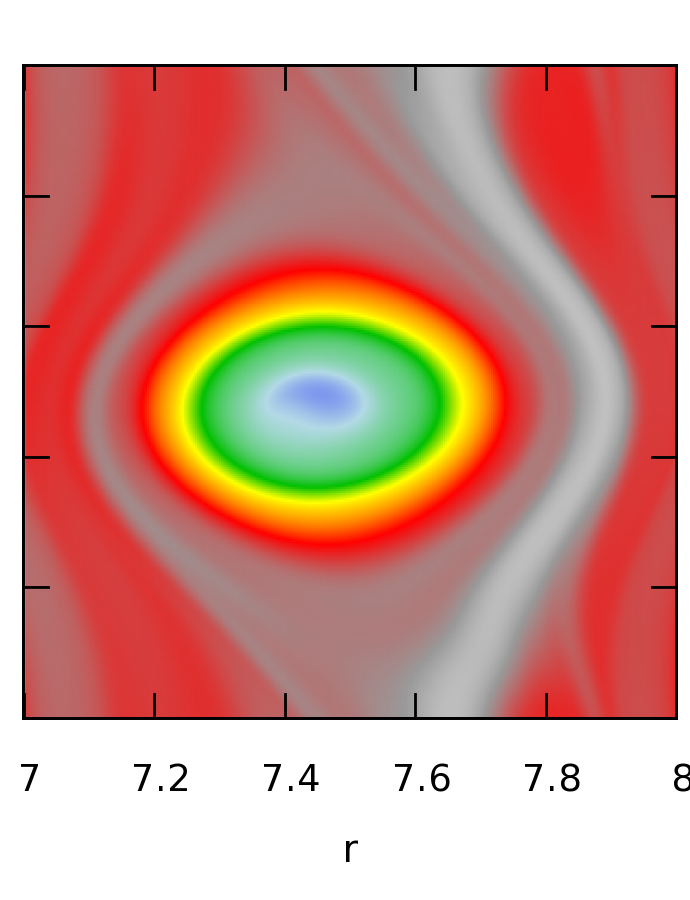}\\
\includegraphics[width=0.5cm]{decalage.png}
\includegraphics[width=0.37cm]{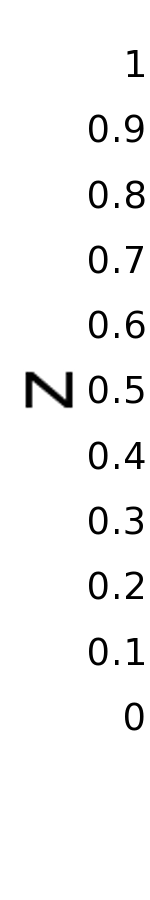}
\includegraphics[width=3.2cm]{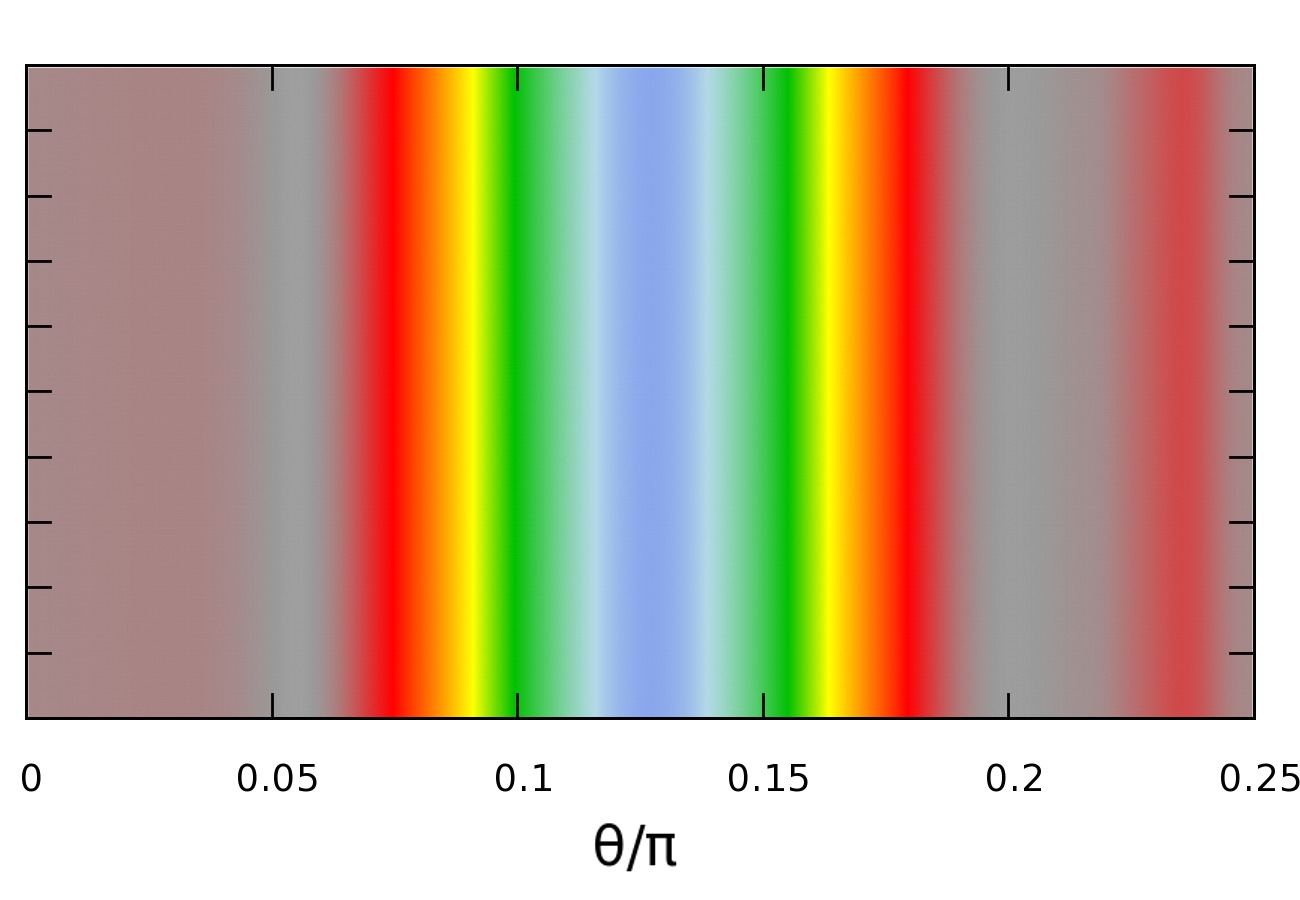}
\includegraphics[width=3.2cm]{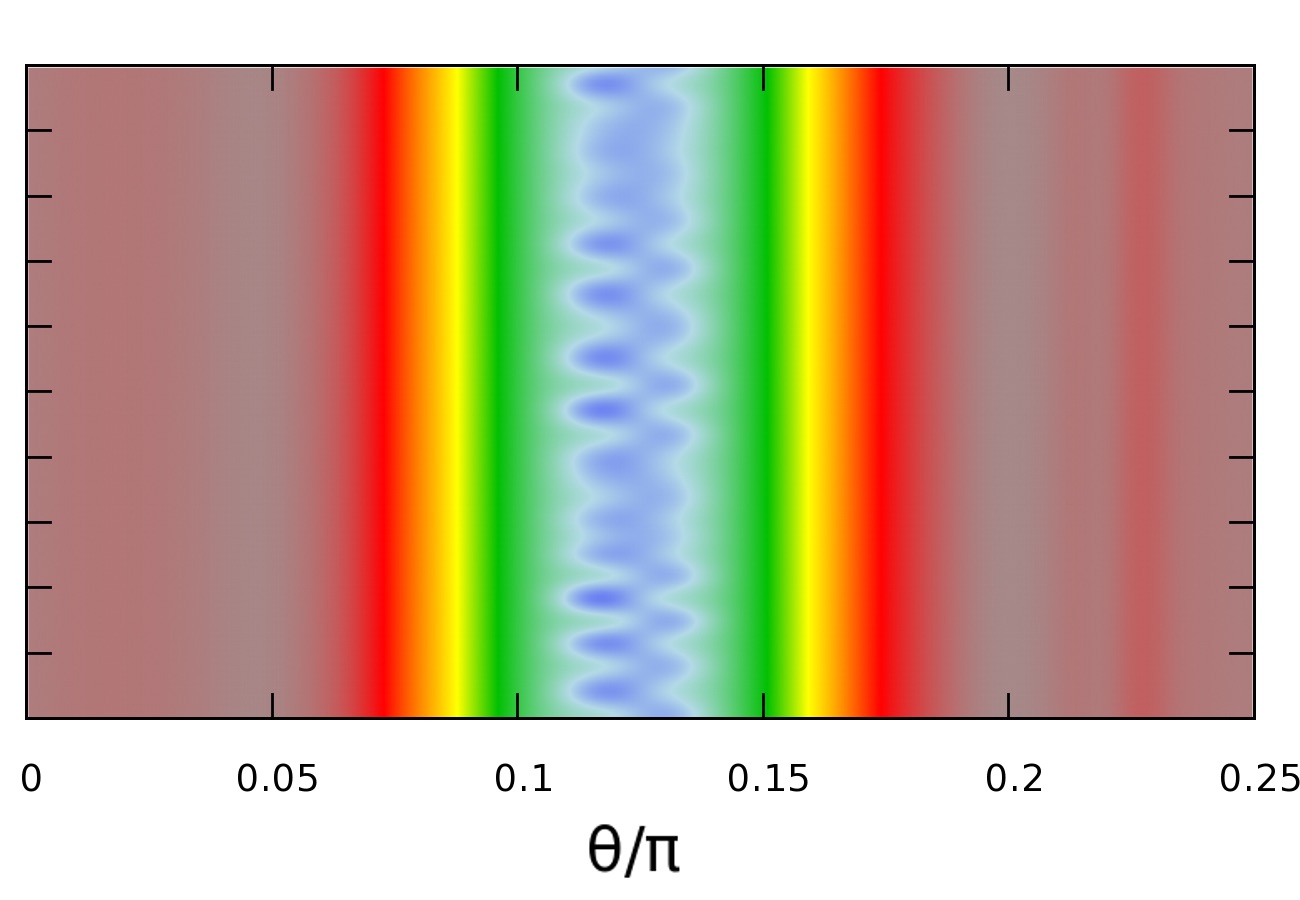}
\includegraphics[width=3.2cm]{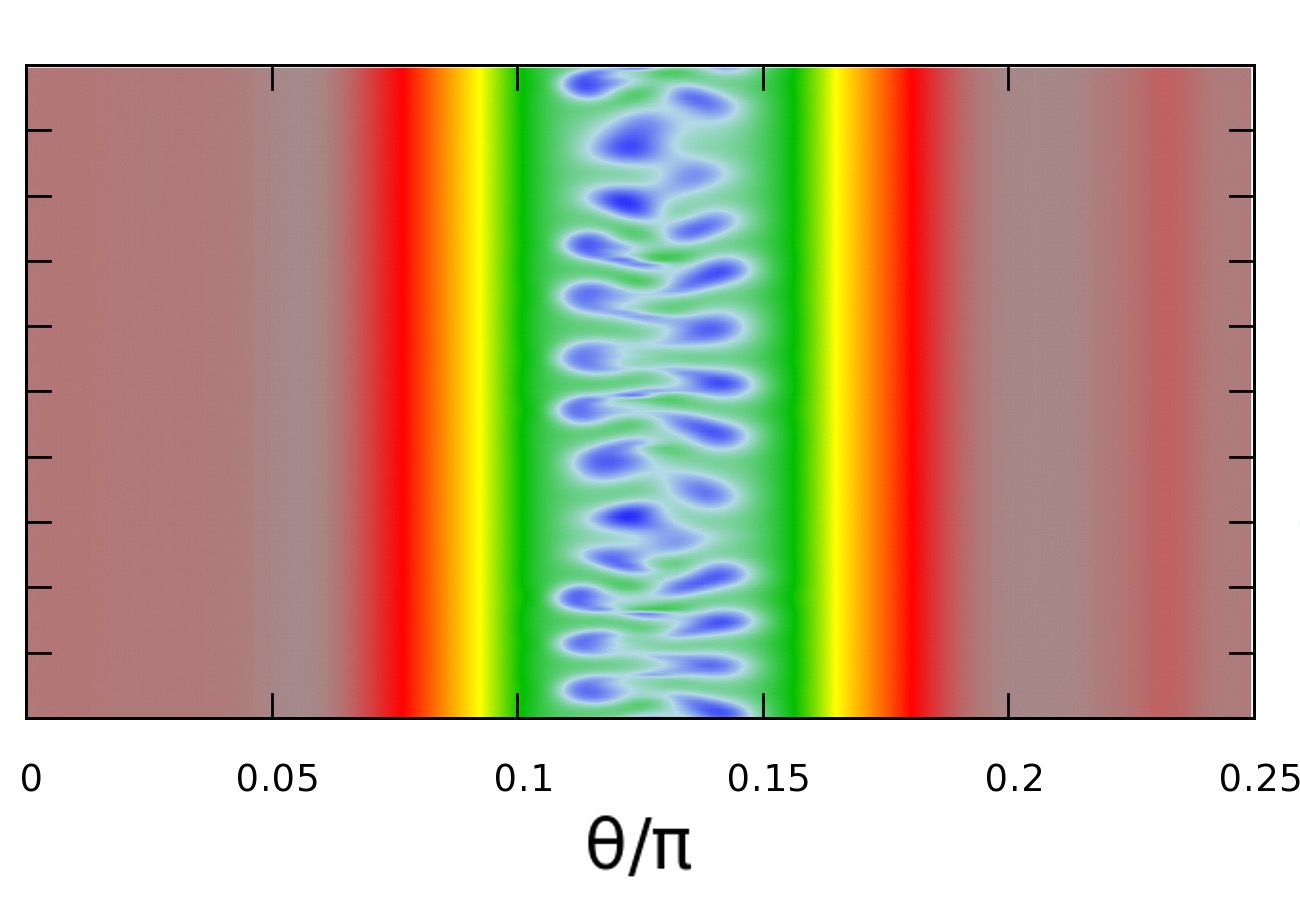}
\includegraphics[width=3.2cm]{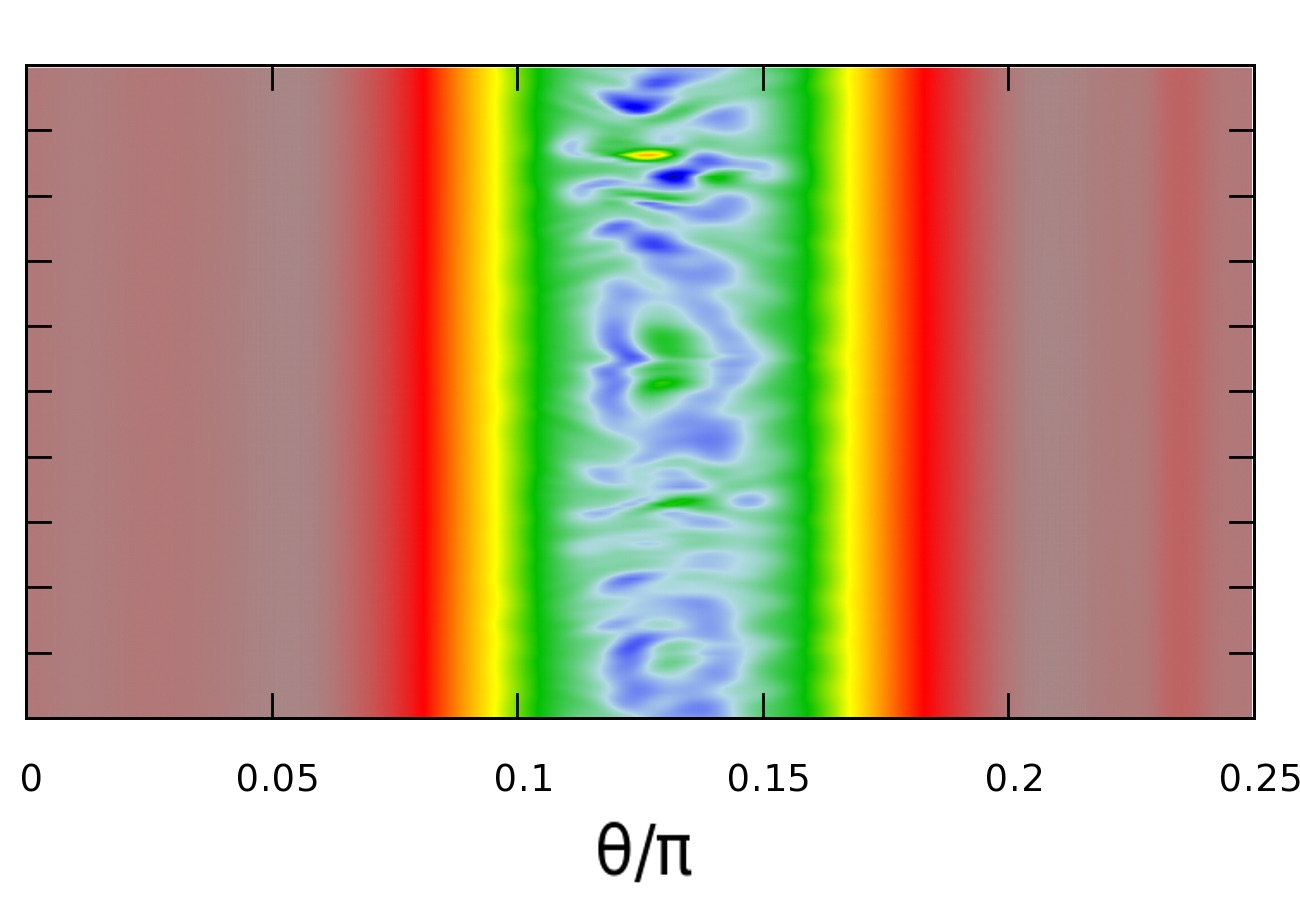}
\includegraphics[width=3.2cm]{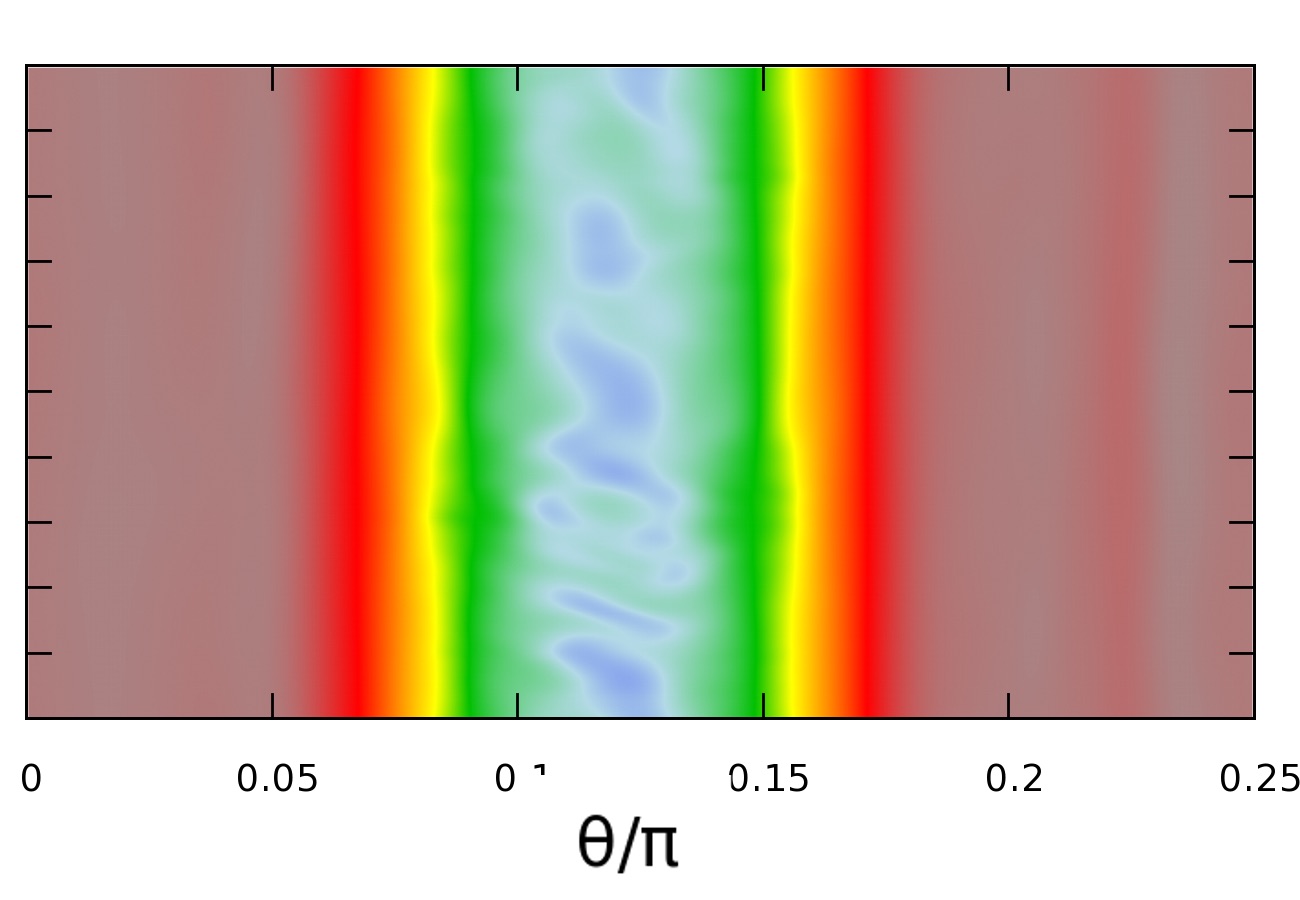}
\end{minipage}
\begin{minipage}[c]{.03\linewidth}
\includegraphics[width=1.3cm, height=7cm]{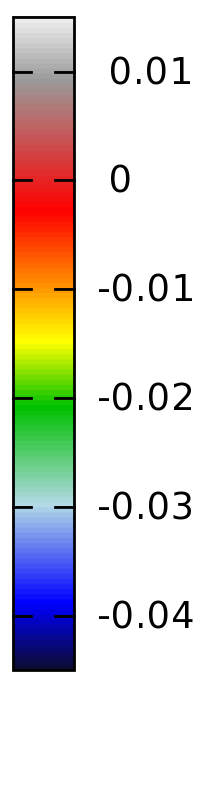}
\end{minipage}
\caption{Vorticity of a 3D baroclinic vortex with aspect ratio $\chi=4$: at $t=0$ and after 20, 22, 24 and 40 rotations; in a 3D perspective (top), in the $r$-$\theta$ plane (middle) and in the $\theta-z$ plane (bottom).}
\label{ibie}%
\end{figure*}

The continuous growth of the vortices by baroclinic amplification obviously raises the question of their evolution when their aspect-ratio reaches a value small enough to be affected by the elliptic instability (Lesur and Papaloizou, 2009). In the case of unstratified disks, LP10 found that vortices can survive this 3D instability if the baroclinic production of vorticity  balances the mixing induced by the elliptical instability.
Here, the problem is revisited in the case of compressible flow. 
In order to avoid the long (time consuming) integration necessary to produce a single and strong enough vortex we started the simulations from an approximate vortex solution. This solution is build up in a two steps procedure:  first, a 2D baroclinic vortex is formed using the 2D version of the code and we let it grow until it reaches an aspect ratio $\chi=4$; then, the 2D vortex is considered in a 3D setting by piling up the vortex profile in the third direction. The resulting vortex is a columnar vortex in an unstratified 3D disk. 
Figure \ref{ibie} shows the evolution of a 3D baroclinic vortex constructed in such a way. It is clearly affected by the elliptical instability with some vertical wavy motions but the instability is never strong enough to completely destroy the vortex. We observe that the vortex re-organizes in a vortical structure with a turbulent core that can survive till the end of the simulation. The results we found are consistent with those of \citet{Lesur2010} and  \citet{Lyra2011} who found the emergence of vertical motions inside the vortices when their aspect ratio satisfies $\chi<4$.

\subsection{Vertically stratified disk}
We now consider a fully 3D stratified disk.\
\subsubsection{Radial stratification and layering}
The disk structure is governed by a vertical equilibrium between gravity and pressure gradient
\begin{equation}
{ \frac{\partial P}{\partial z}}=-\rho { \frac z{(r^2+z^2)^{3/2}}}  ~~~.
\end{equation}
As in the two-dimensional case, the radial profiles of the temperature and the surface density are assumed to be simple power laws. It is then easy to derive the gas density\\
\begin{equation}
\rho_e=r^{-\beta} \exp \left({ \frac 1{T}} \left({ \frac 1 {\sqrt{r^2+z^2}}} - { \frac 1{r}} \right) \right)   \approx r^{-\beta} e^{ -{\frac1{2rT}} ({ \frac z{r} })^2  }  
\label{rho3D}
\end{equation}
in which $\beta=p+(3-q)/2$ depends on the two indices $p$ and $q$. Radial stratification is characterized by the Brunt-V\"{a}is\"{a}l\"{a} frequency\\
\begin{equation} 
N_r^2 = -{ \frac{RT}\gamma}  \left( { \frac{ \partial ~ln\rho_e}{\partial r} } - { \frac q{r} }\right)\left( (1-\gamma){ \frac{ \partial ~ln\rho_e}{\partial r} } - {   \frac q{r} }\right)  
\end{equation}
which in the thin disk approximation ($z/r \ll1$) and using (\ref{exp:HD}) leads to\\
\begin{multline}
N_r^2 \approx  \gamma(\gamma-1)  \left({   \frac{3-q}2 }\right)^2  { \frac{RT}{r}}  \left( { z^2} - {z_2^2} \right)  \left( { z^2} - {z_1^2} \right)   
\end{multline}
where
\begin{equation} 
z_1= H(r)\sqrt{ { \frac 2{\gamma(\gamma-1)} } { \frac {\beta(\gamma-1)-q }{3-q}} }, 
\end{equation}
\begin{equation} 
z_2= H(r)\sqrt{ { \frac 2{\gamma}} {  \frac{\beta+q}{3-q}}  } .
\end{equation}
So, Schwarchild criterion leads to four possible roots ($\pm z_1$, $\pm z_2$) that indicate the altitudes in the disk where gas stability is changing. \\
In the case of protoplanetary disks,  $q<3$ is a reasonable assumption; therefore, $z_2$ is always a real root but $z_1$ may be imaginary or real, depending on the sign of $\beta(\gamma-1)-q$.\\\
- If  $\beta(\gamma-1)-q<0$  , $z_1$ is imaginary:  stratification is unstable in a mid-plane layer for $z<|z_2|$ and stable in the upper layer for $z>|z_2|$.\\
- If  $\beta(\gamma-1)-q>0$  , $z_1$ is real: stratification is stable, in both the mid-plane layer and the upper layer, but it is  unstable in the intermediate layer for $|z_1|<z<|z_2|$. 

This layering of the stability regions in a three-dimensional disk contrasts with the uniform distribution of stability in a two-dimensional disk in which $N_r^2$ keeps the same sign at any distance from the star. In the following, 
we restrict the analysis to the first case ($\beta(\gamma-1)-q<0$) for which the disk is unstable in a mid-plane layer. 

We set  the power law indices for the density and the temperature to: $\beta=1$ and $q=2$, respectively. 
For these values, we get $|z_2|= 2H$. The disk is then unstable for $z<2H$ and stable for $z> 2H$. 
The computational domain is chosen to include both layers. It is defined in the non-dimensional variables as $7<r<8$, $0<\theta<\pi/4$ and $0<z<1$ (that is $0<z< 2.90H$ at $r=7.5$). The numerical resolution we used in this case is  450$\times$300$\times$300, that is 155$\times$18$\times$100 cells by scale height in the radial, the azimuthal and the vertical directions, respectively.

As for the 2D simulations, the baroclinic instability is initiated by adding perturbations of sufficiently large amplitude to the equilibrium state. 

\subsubsection{Development of the instability}

In a three-dimensional disk the baroclinic instability is found to develop in nearly the same way as in a two-dimensional disk. The first stage corresponding to the formation of small vortices from the density perturbations initially seeded in the background is recovered. These vortices mainly appear in the mid-plane of the disk like in the 2D case but they also possess a small vertical extent. Then, once formed,  they tend to grow slowly in size and strength.

\begin{figure}[!h]
\centering
\includegraphics[width=9cm]{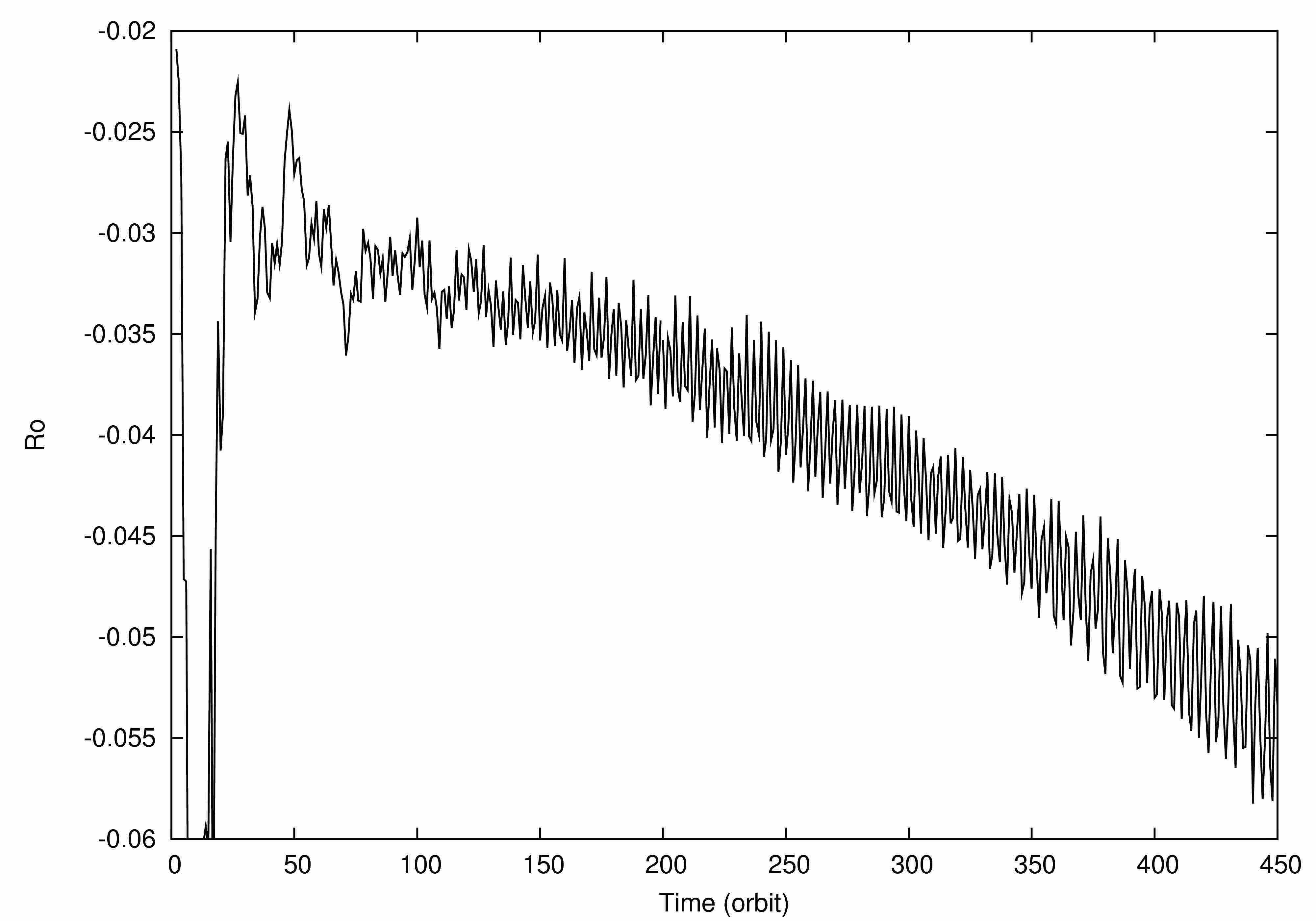}
\caption{Rossby number as a function of time of the strongest vortex formed by baroclinic instability in a 3D disk.}
\label{ib3De}
\end{figure}

Figure \ref{ib3De} shows the evolution of the Rossby number of the strongest vortex formed at the end of the first stage. It clearly illustrates the vorticity amplification process occurring during the formation and growth of the vortex.

\begin{figure*}[!h]
\begin{minipage}[c]{.9\linewidth}
\includegraphics[width=.2cm, height=3.5cm]{theta.png}
\includegraphics[width=4cm, height=3.5cm]{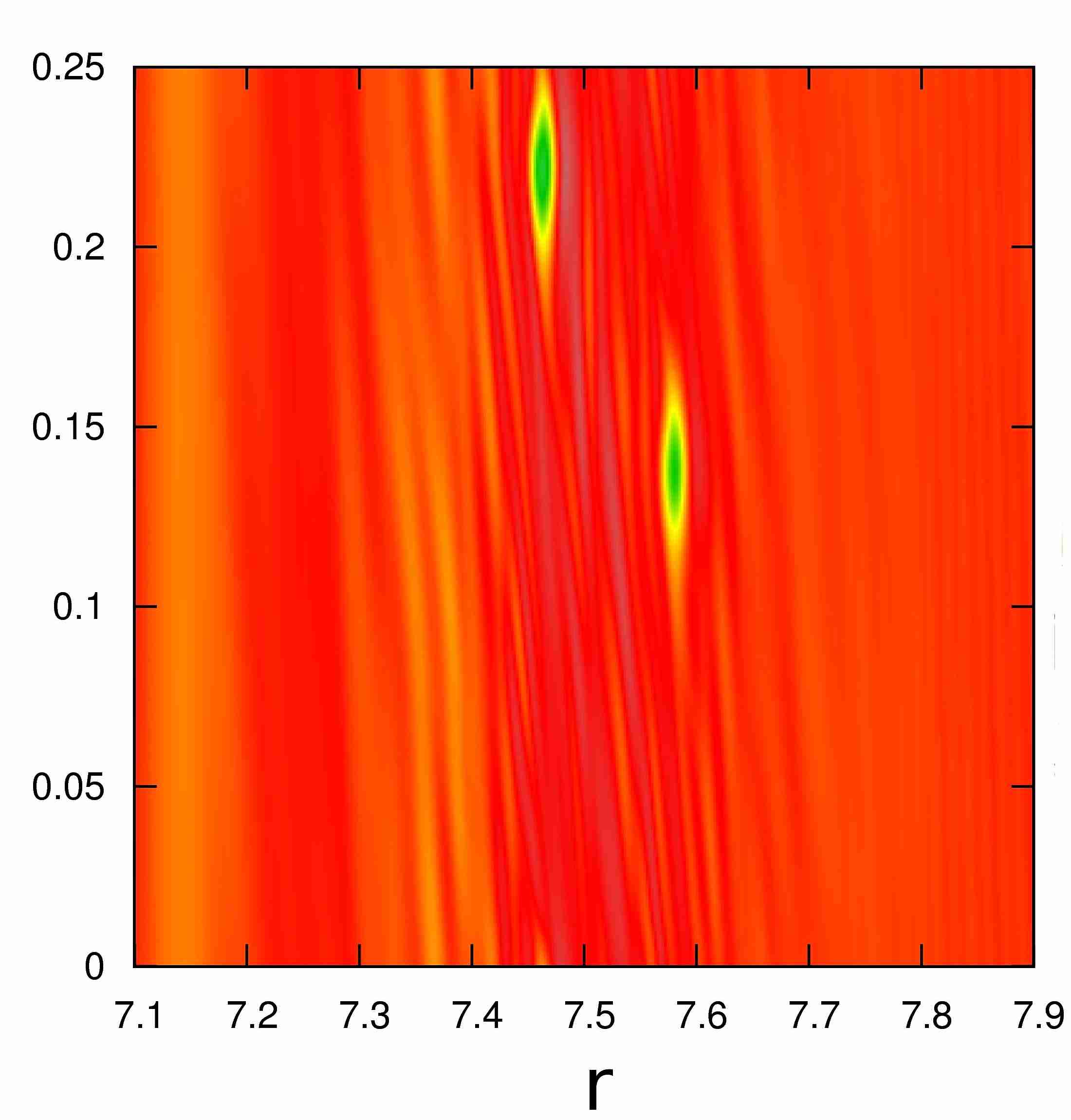}
\includegraphics[width=4cm, height=3.5cm]{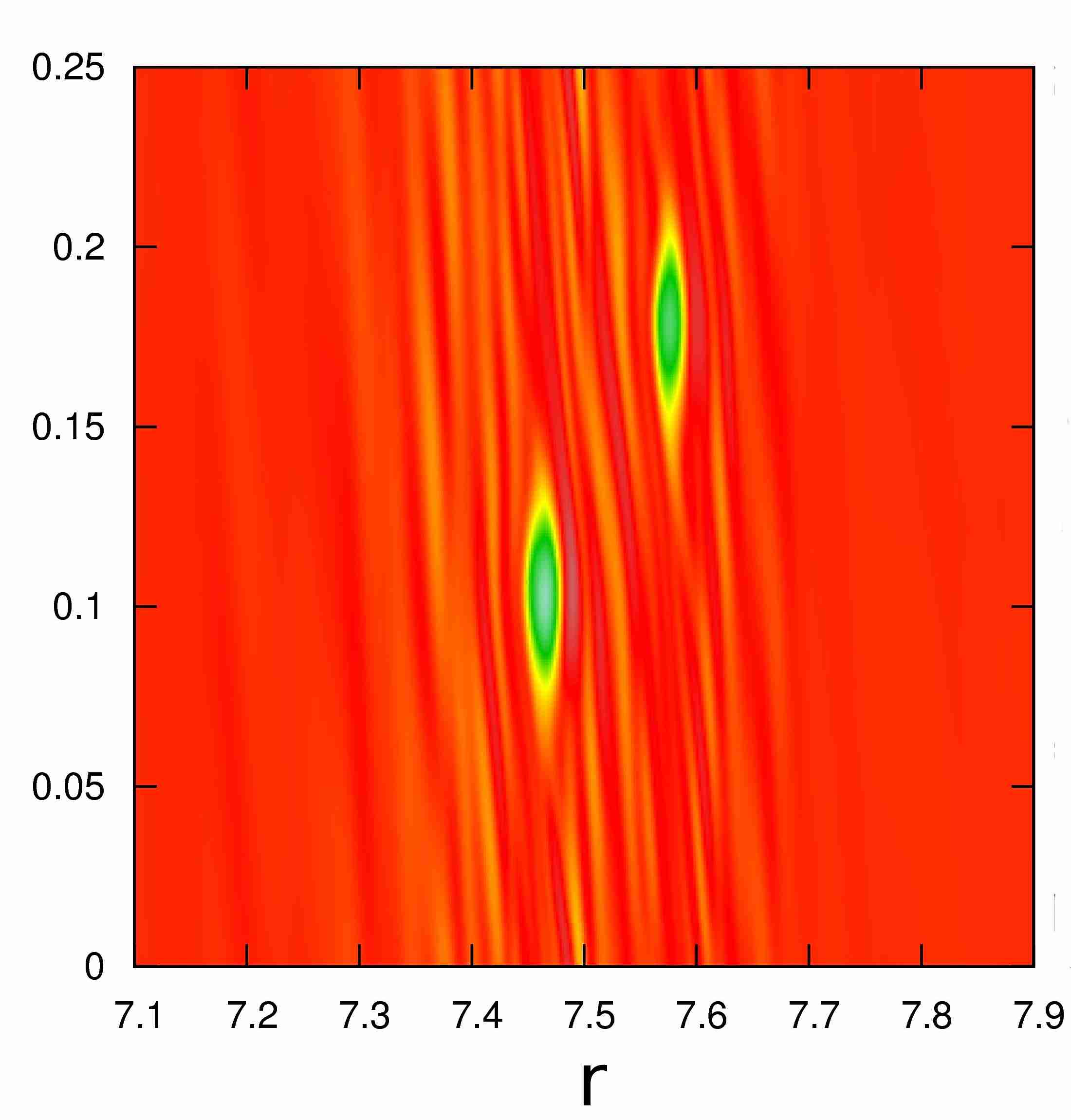}
\includegraphics[width=4cm, height=3.5cm]{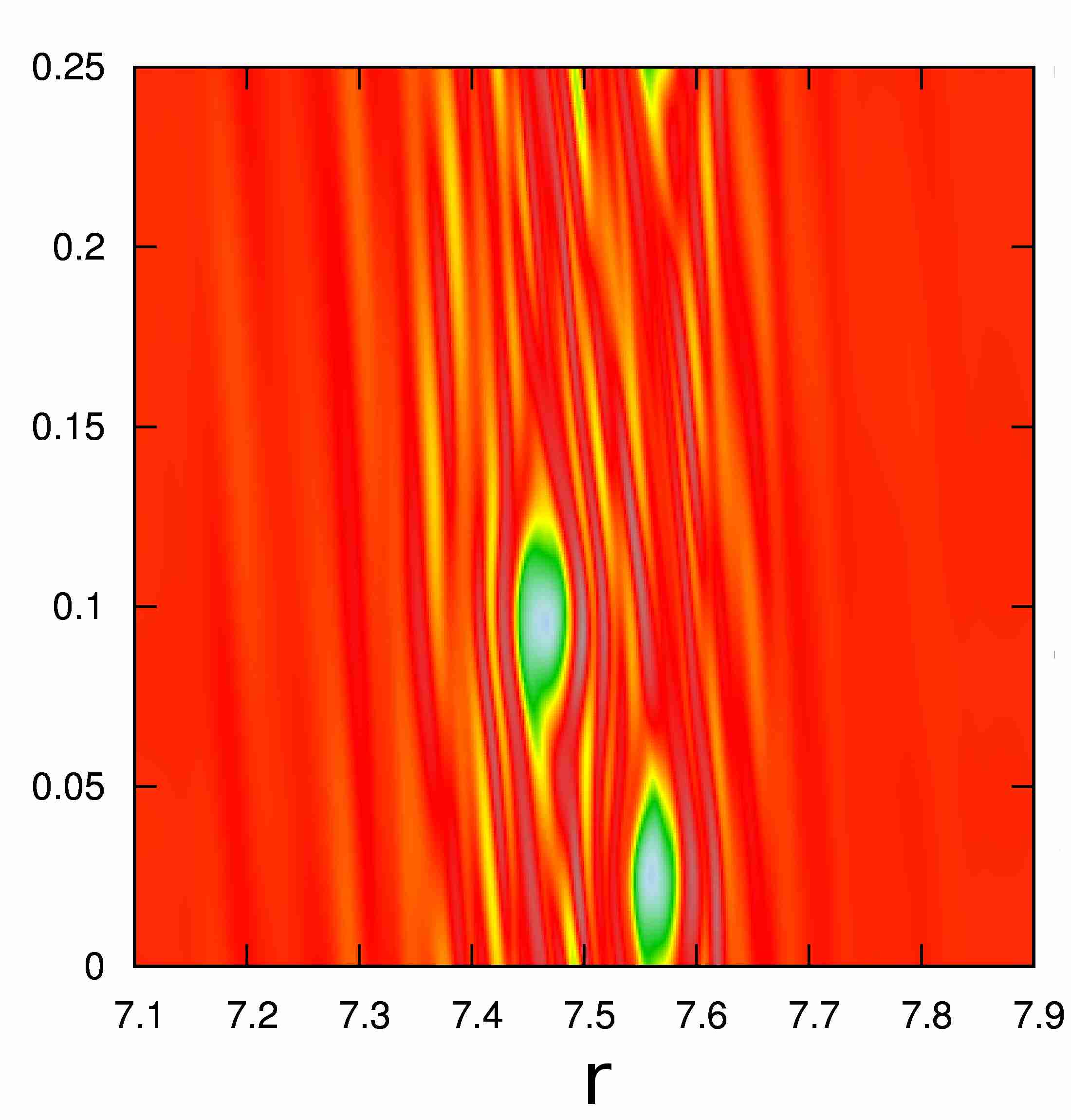}
\includegraphics[width=4cm, height=3.5cm]{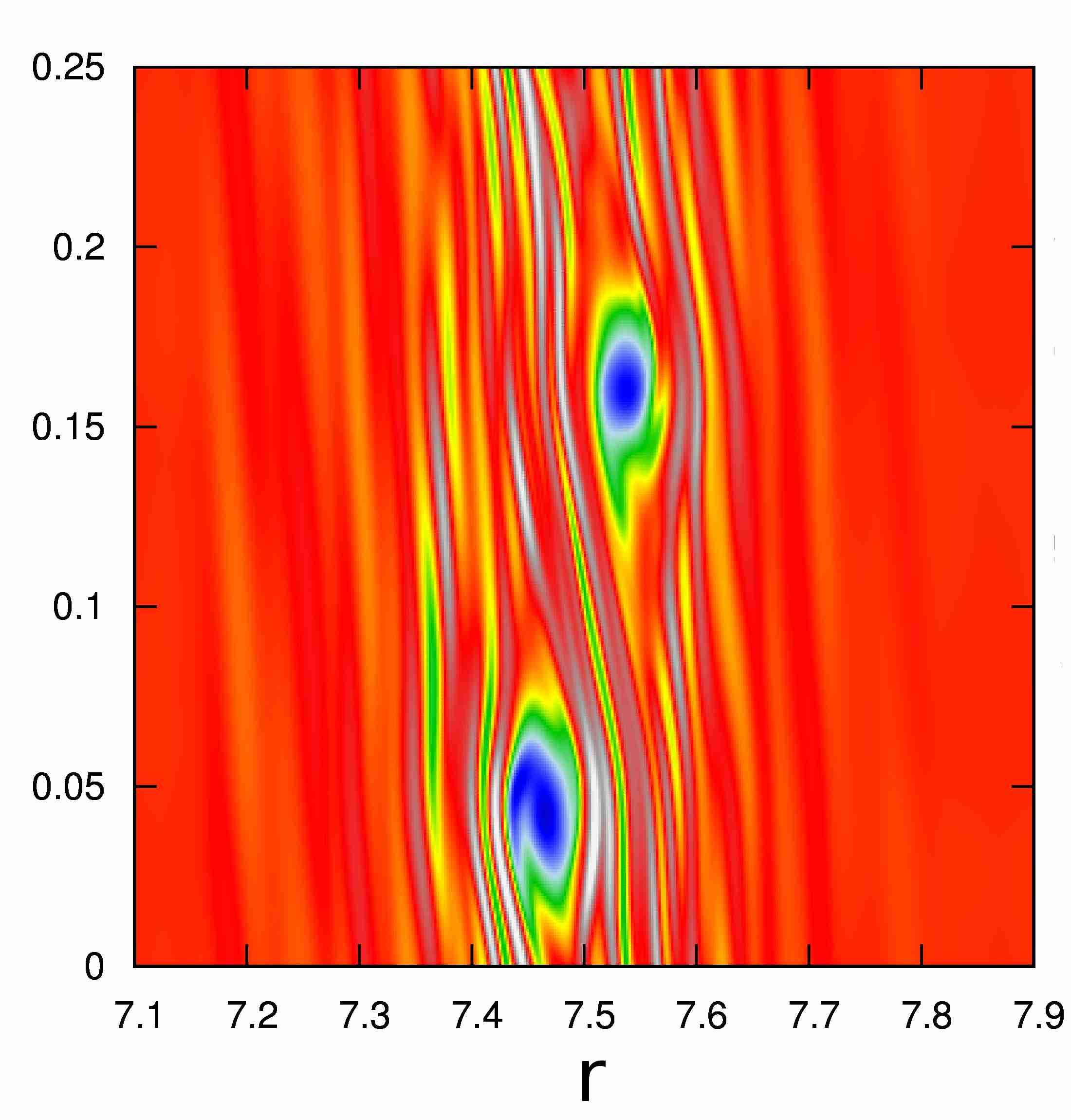}\\
\includegraphics[width=0.2cm, height=3.5cm]{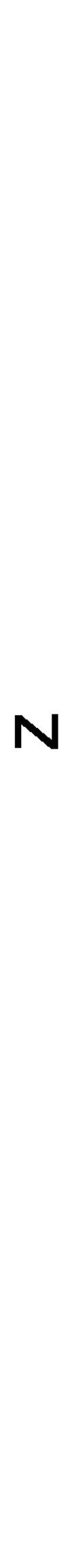}
\includegraphics[width=4cm, height=3.5cm]{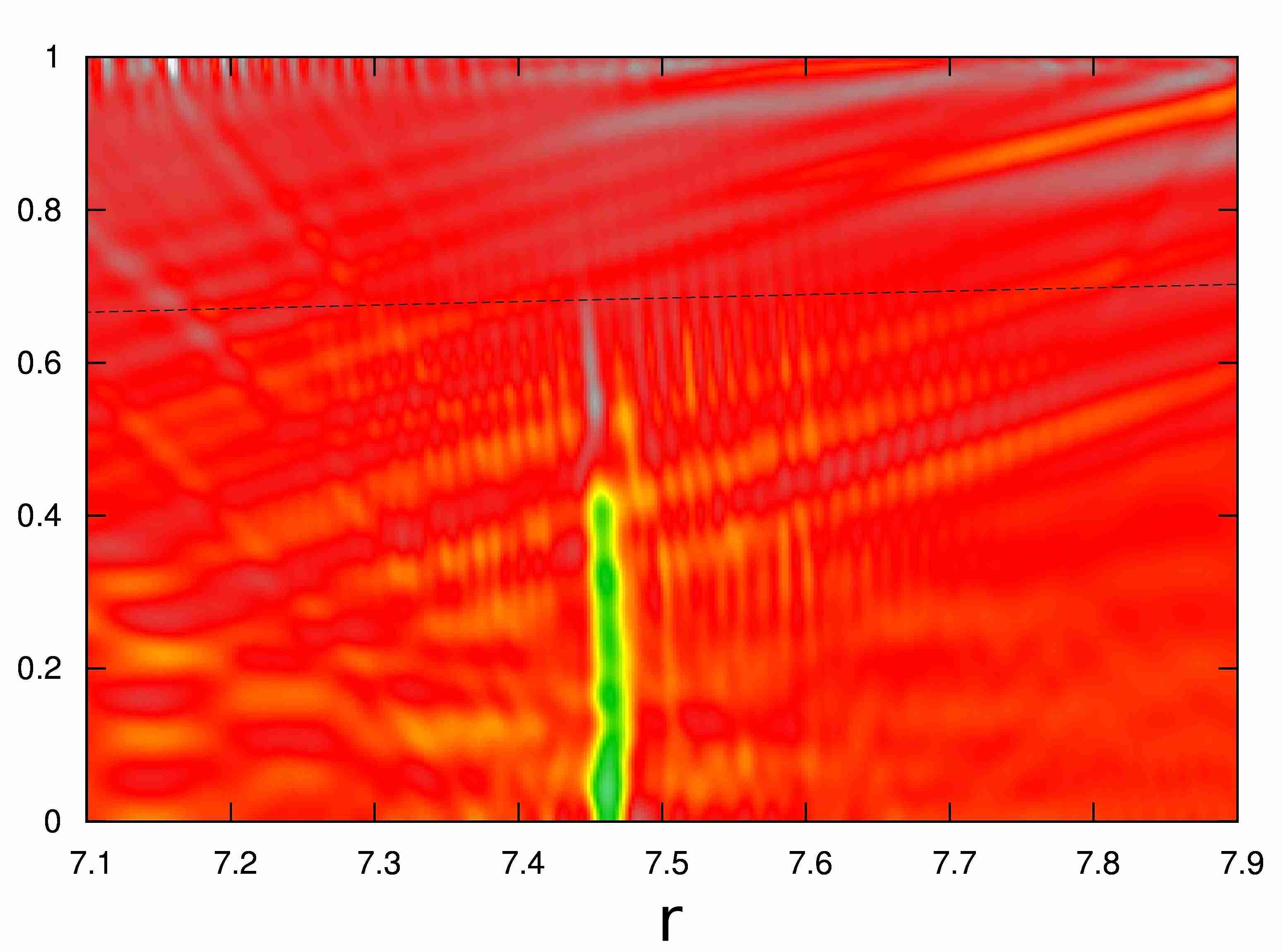}
\includegraphics[width=4cm, height=3.5cm]{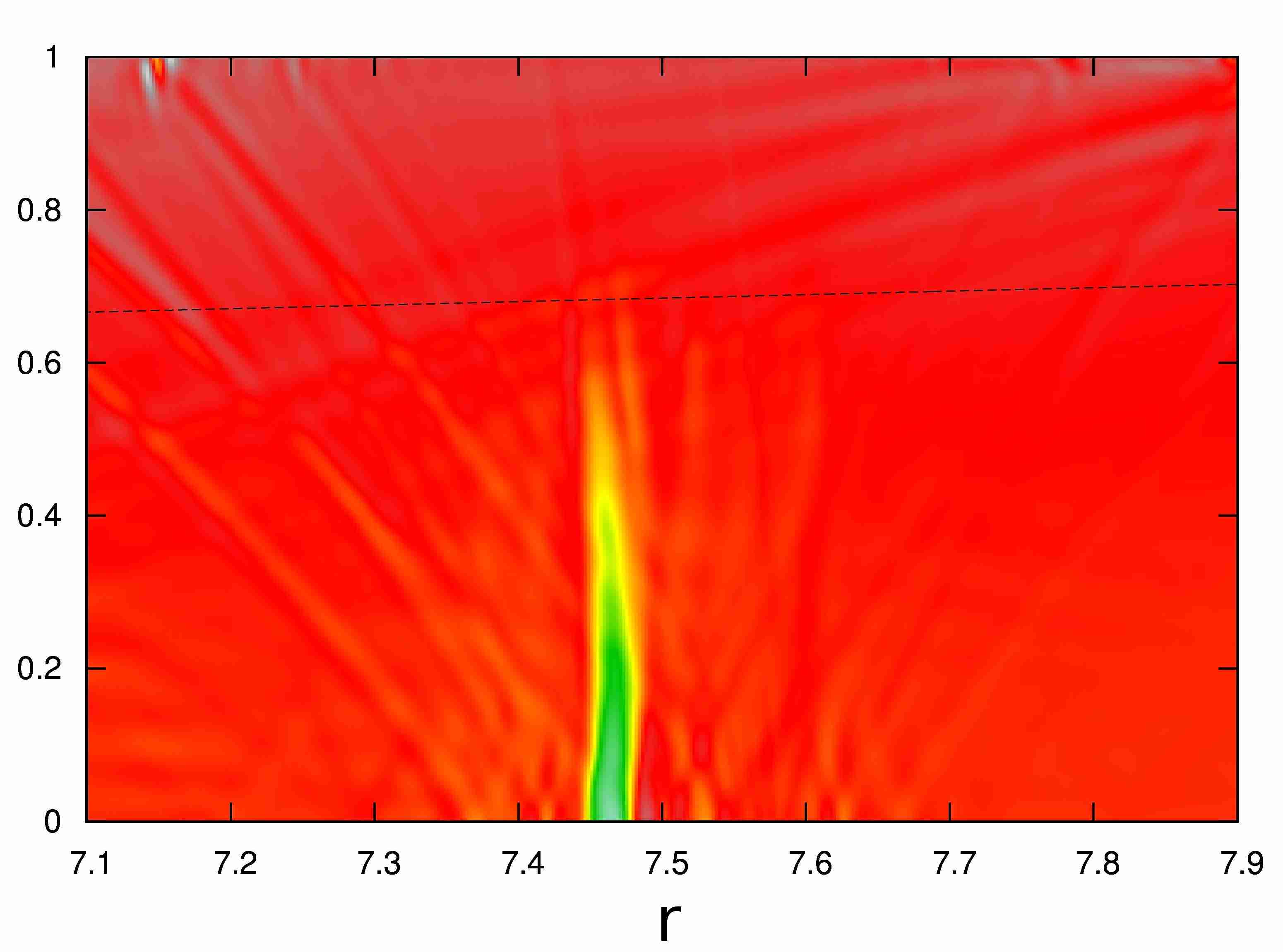}
\includegraphics[width=4cm, height=3.5cm]{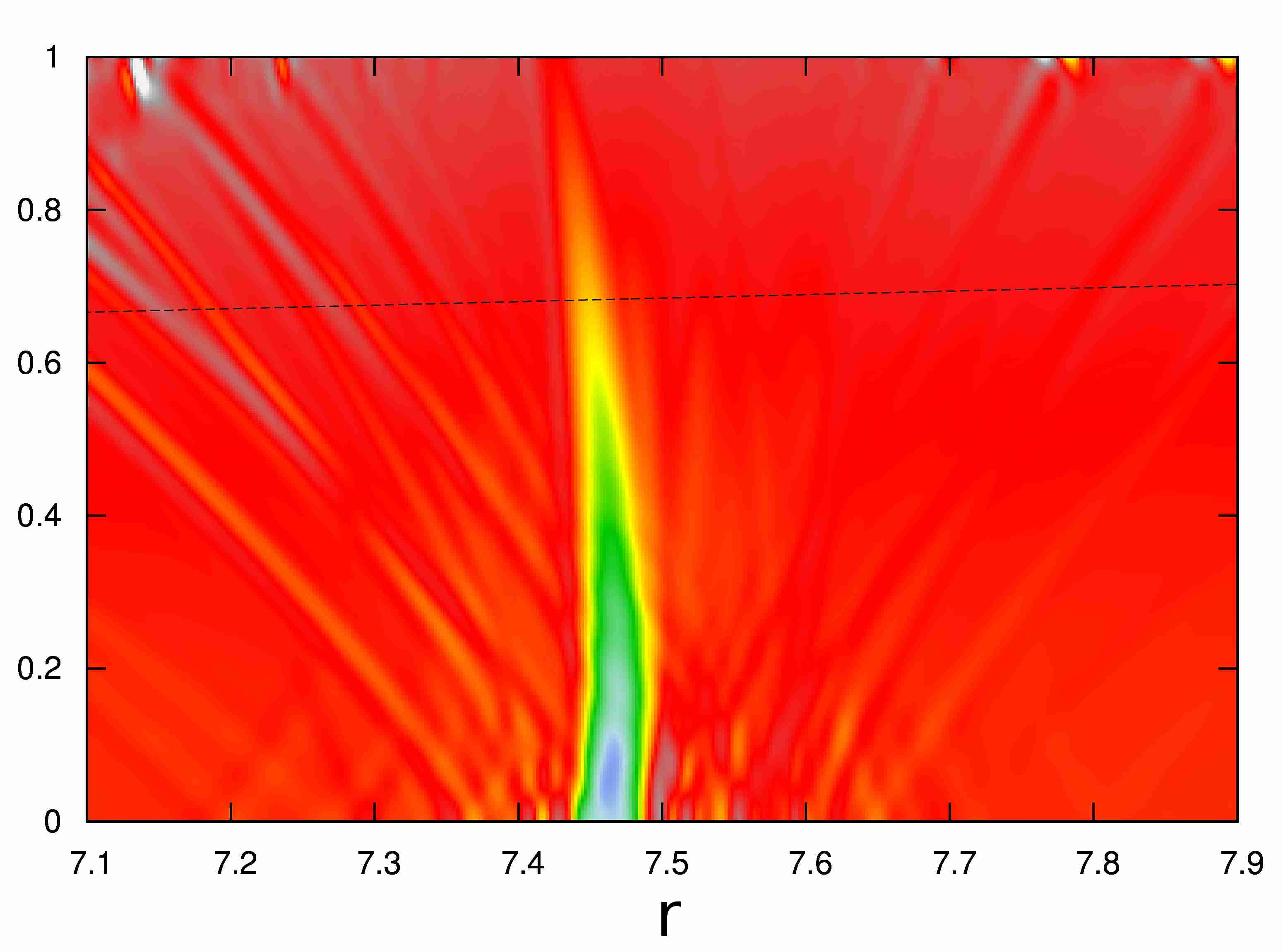}
\includegraphics[width=4cm, height=3.5cm]{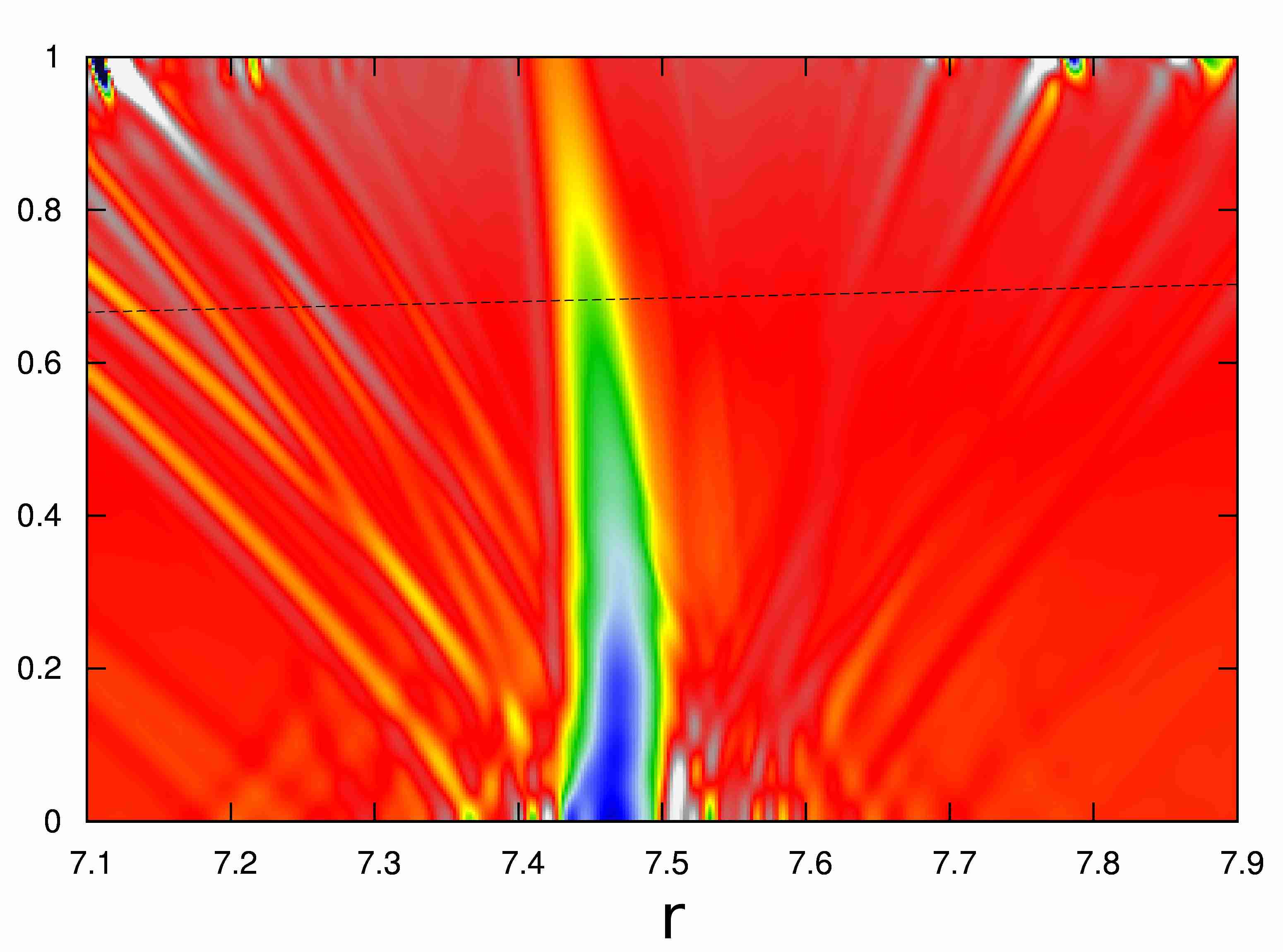}\\
\includegraphics[width=0.2cm, height=3.5cm]{z.jpg}
\includegraphics[width=4cm, height=3.5cm]{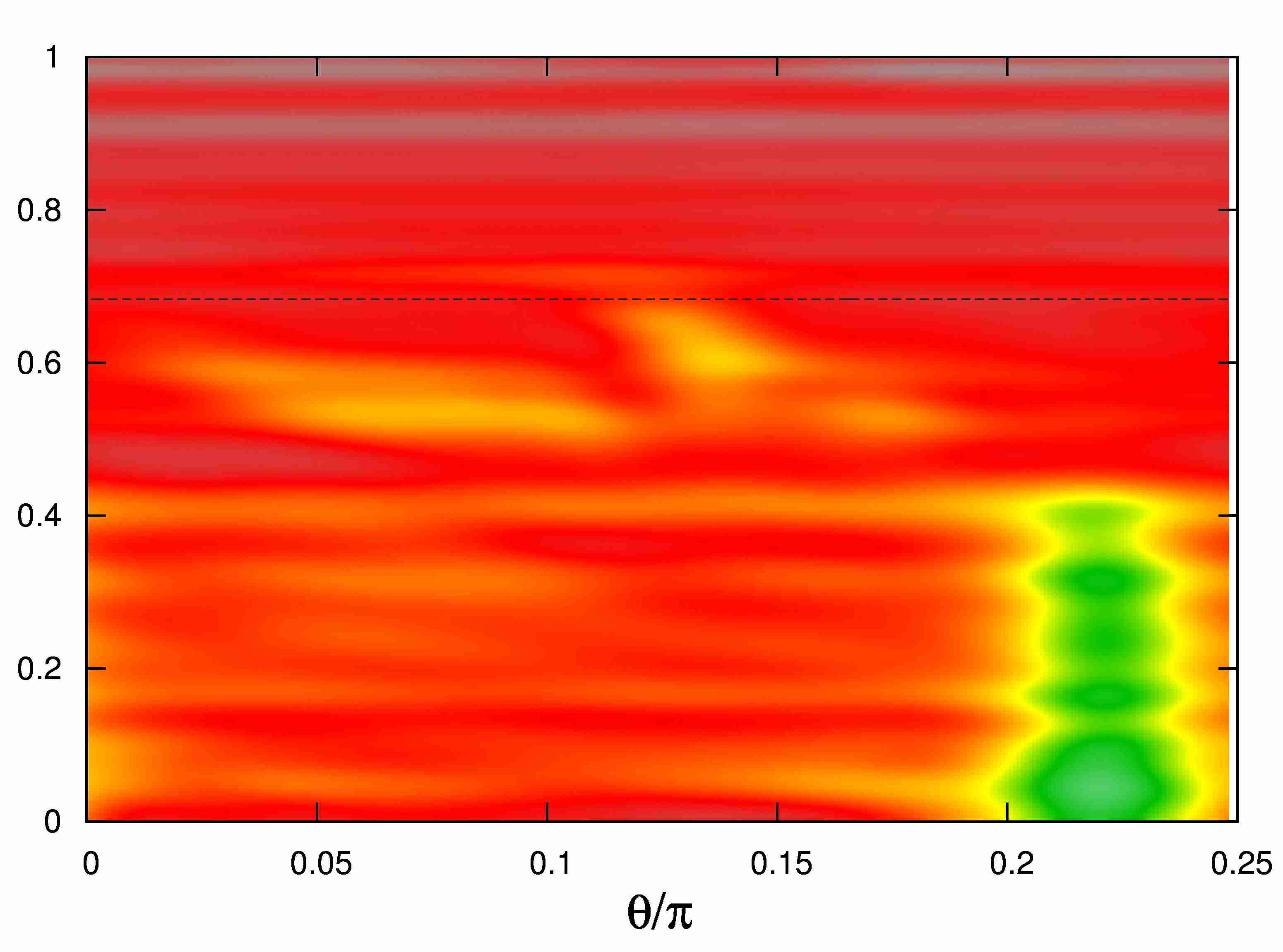}
\includegraphics[width=4cm, height=3.5cm]{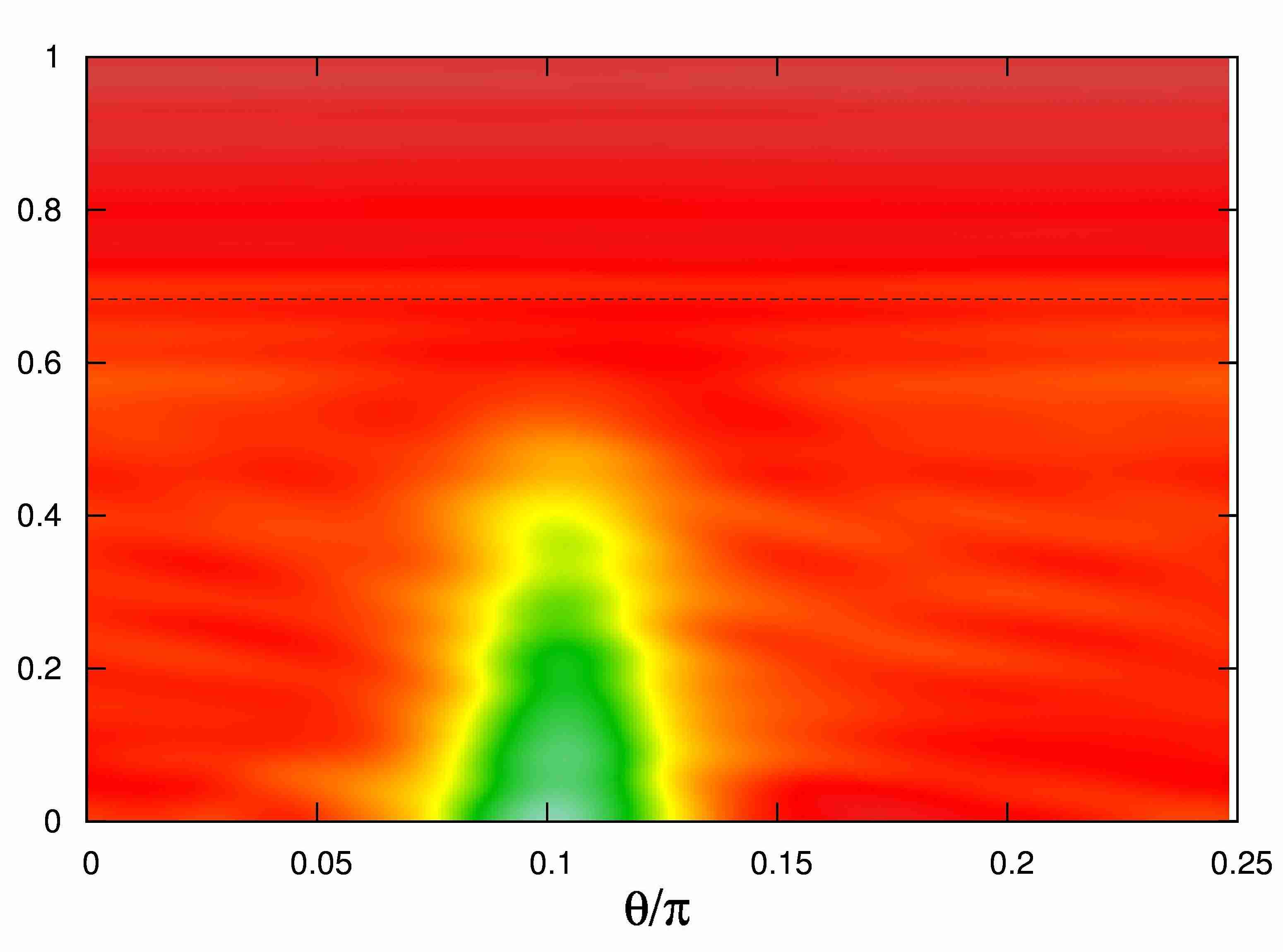}
\includegraphics[width=4cm, height=3.5cm]{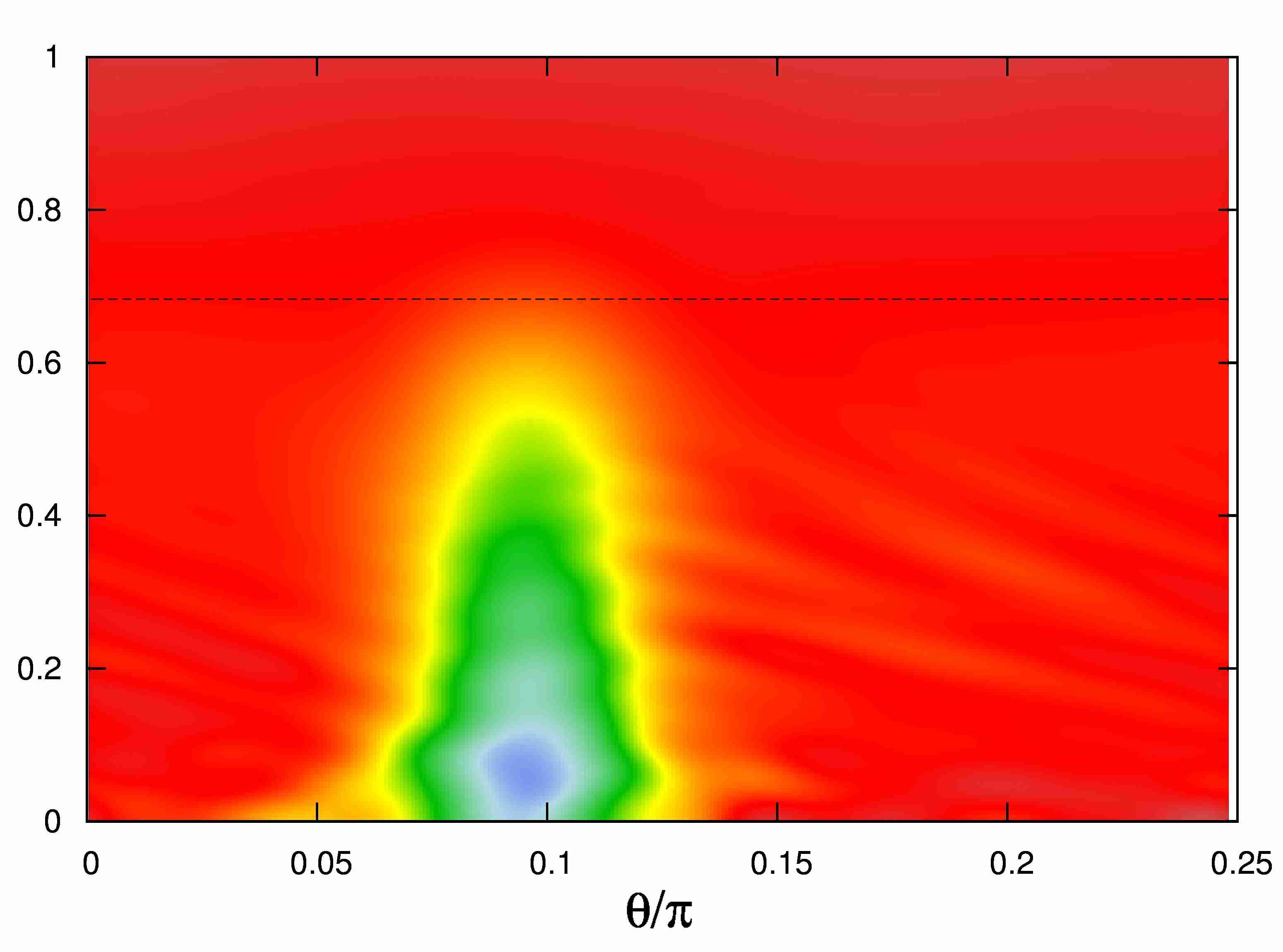}
\includegraphics[width=4cm, height=3.5cm]{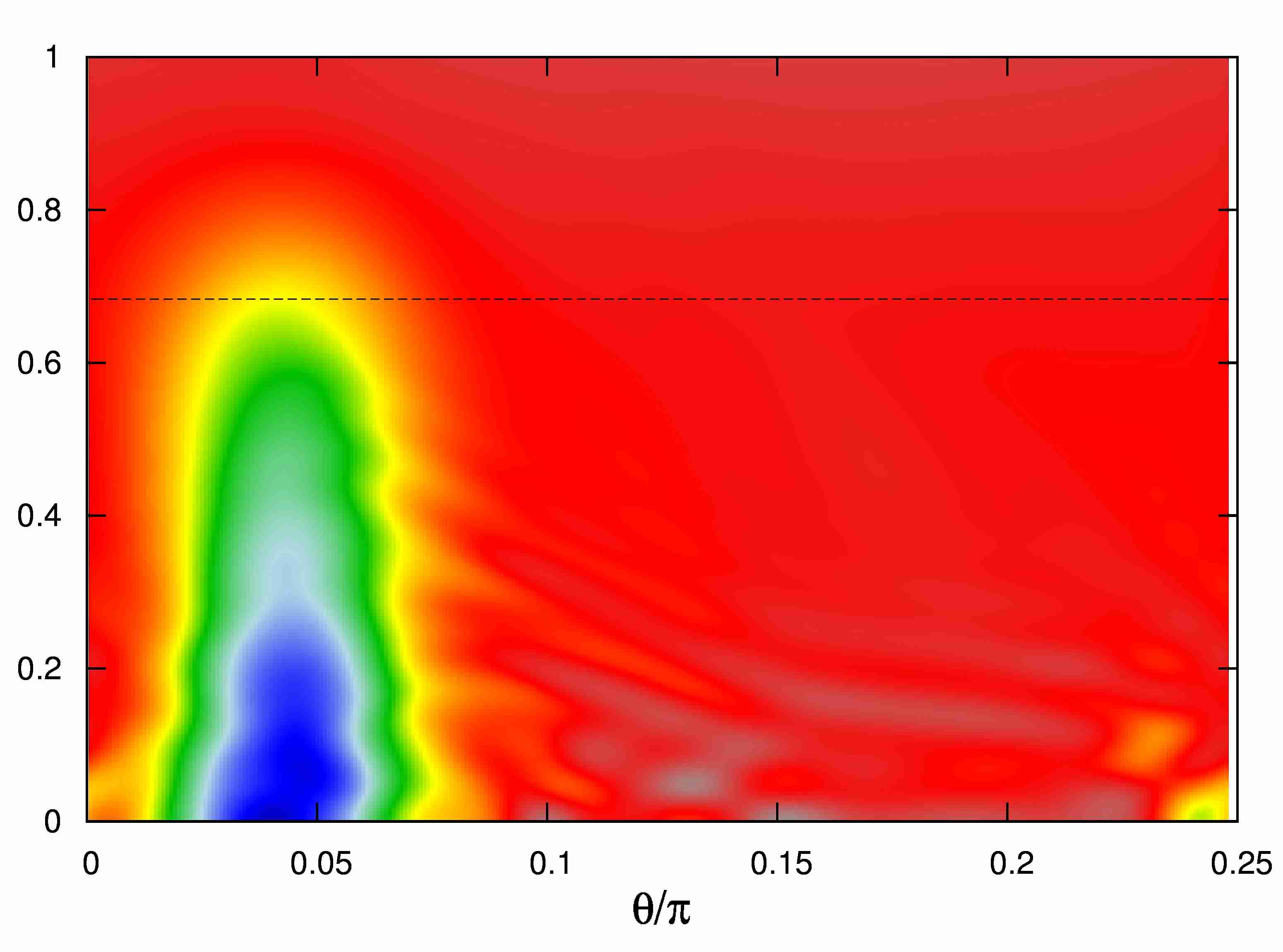}
\end{minipage}
\begin{minipage}[c]{.05\linewidth}
\includegraphics[width=1cm, height=7cm]{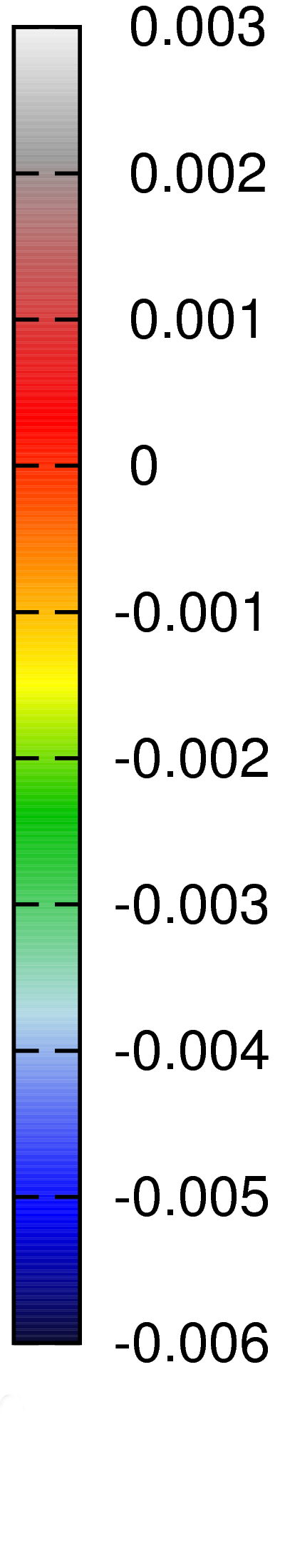}
\end{minipage}
\caption{Vorticity map after 50, 250, 400 et 450 rotations, in the midplane (top), and in cuts through the strongest vortex in the r-z plane (middle) and in the $\theta$-z plane  (bottom). The dashed line shows the height where $N_r^2$ become positive.}
\label{ib3Dplot}
\end{figure*}

Figure \ref{ib3Dplot} shows how vortices  grow and also how they expend in the vertical direction. We observe the formation of slightly tilted columns of vorticity that progressively fill the whole disk thickness, including the stable upper layer. This spread of the vorticity in the stable layers of the disk is an interesting and new result. The aspect ratio of the vortices decreases as the vorticity increases, and reaches 10 at the end of the simulation. This large aspect ratio makes the vortex relatively stable with respect to the elliptical instability.
 
\subsubsection{Formation of vortex layers }

\begin{figure*}[lh]
\centering
\includegraphics[width=12cm, height=6.75cm]{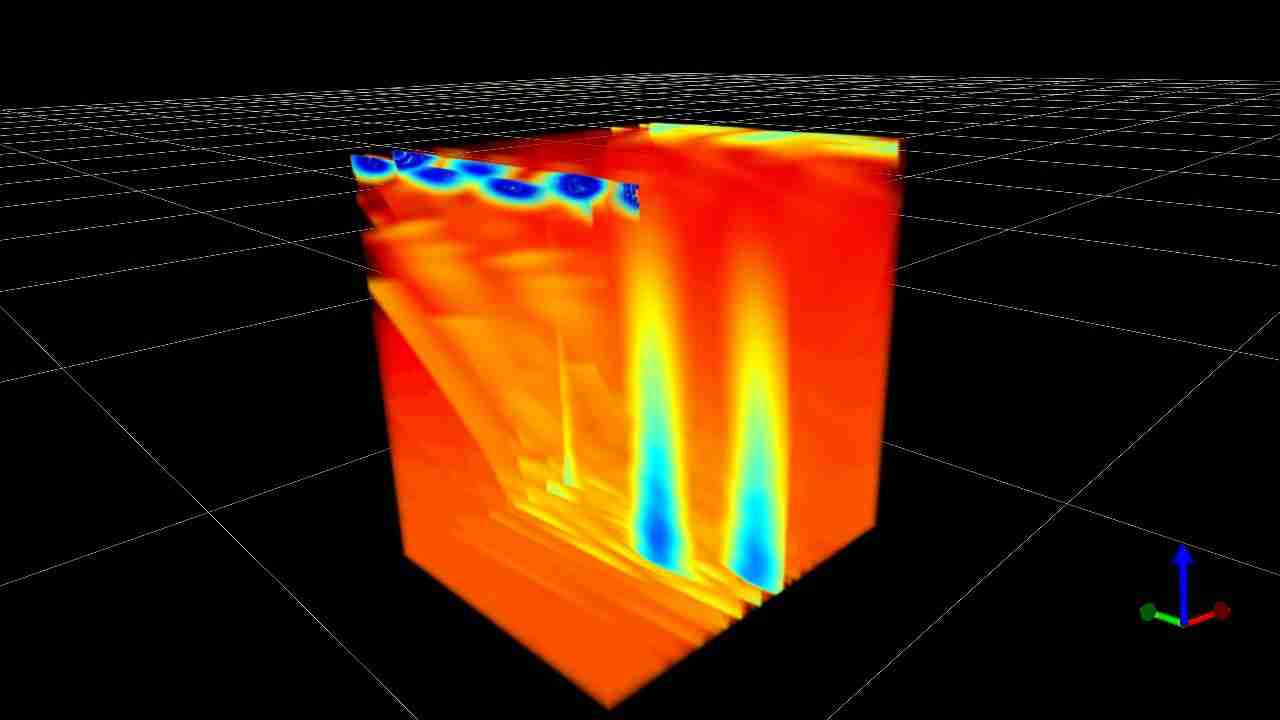}\\
\includegraphics[width=12cm, height=6.75cm]{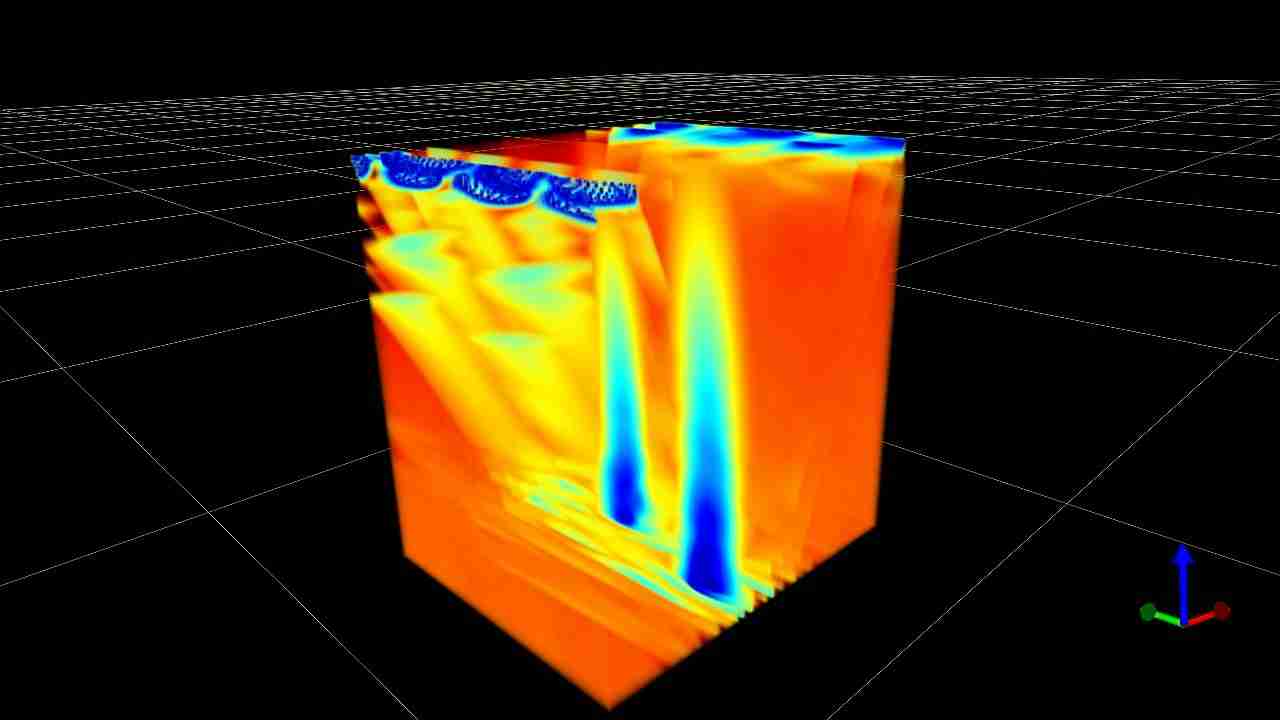}\\
\caption{ Three-dimensional distribution of the vorticity in a fully stratified disk, after 250, 400 and 445 rotations (from top to bottom, respectively). The colored arrows correspond to the radial (red), the azimuthal (green) and to the vertical (blue) directions. }
\label{zombie-3D}
\end{figure*}

In this section, we focus on the weaker features observed on Fig.~\ref{ib3Dplot}  after a few hundred rotations. 
We clearly see  on the vorticity contours in ($r$-$z$) plane (middle row of Fig.~\ref{ib3Dplot}) the presence of linear rays
whose amplitude increases in time, especially in the upper-disk region. 
These rays  exhibit an azimuthal structure which is visible on the 3D plots shown in Fig.~\ref{zombie-3D}.  According to the inclination of the ray, we observe 2, 3 or 4 vortices, corresponding to the azimuthal wavenumbers  $m = 16$, $24$ or $32$ in our $\pi/4$ 
angular computation domain.  
The amplitude of the vorticity perturbation increases along the rays, and is found to be maximum at the boundary of the computation box,
where it forms well-defined anticyclonic vortices (see Fig.~ \ref{zombie-3D} at 400 rotations) that
tend to merge two by two  (see Fig.~\ref{zombie-3D} at 450 rotations). 

We think that these observations could correspond to the first step of what has been christened the Zombie Vortex Instability (ZVI)  by \citet{Marcus2015}. 
\citet{Barranco2005}  were the first to observe the formation of off-midplane vortices during the destabilization of a mid-plane vortex in a stratified disk.
\citet{Marcus2013_2} attributed the formation of these vortices to the presence of baroclinic critical layers. They further showed that once formed  these vortices 
could force new critical layers that create new vortices and so on by a self-replication process. It is this instability process which has been called ZVI. 
Although \citet{Marcus2015} demonstrated a certain robustness of the process, it has never been observed in a fully compressible simulation of 3D Keplerian disks.

\begin{figure*}[!h]
\centering
\includegraphics[width=9cm, height=8cm]{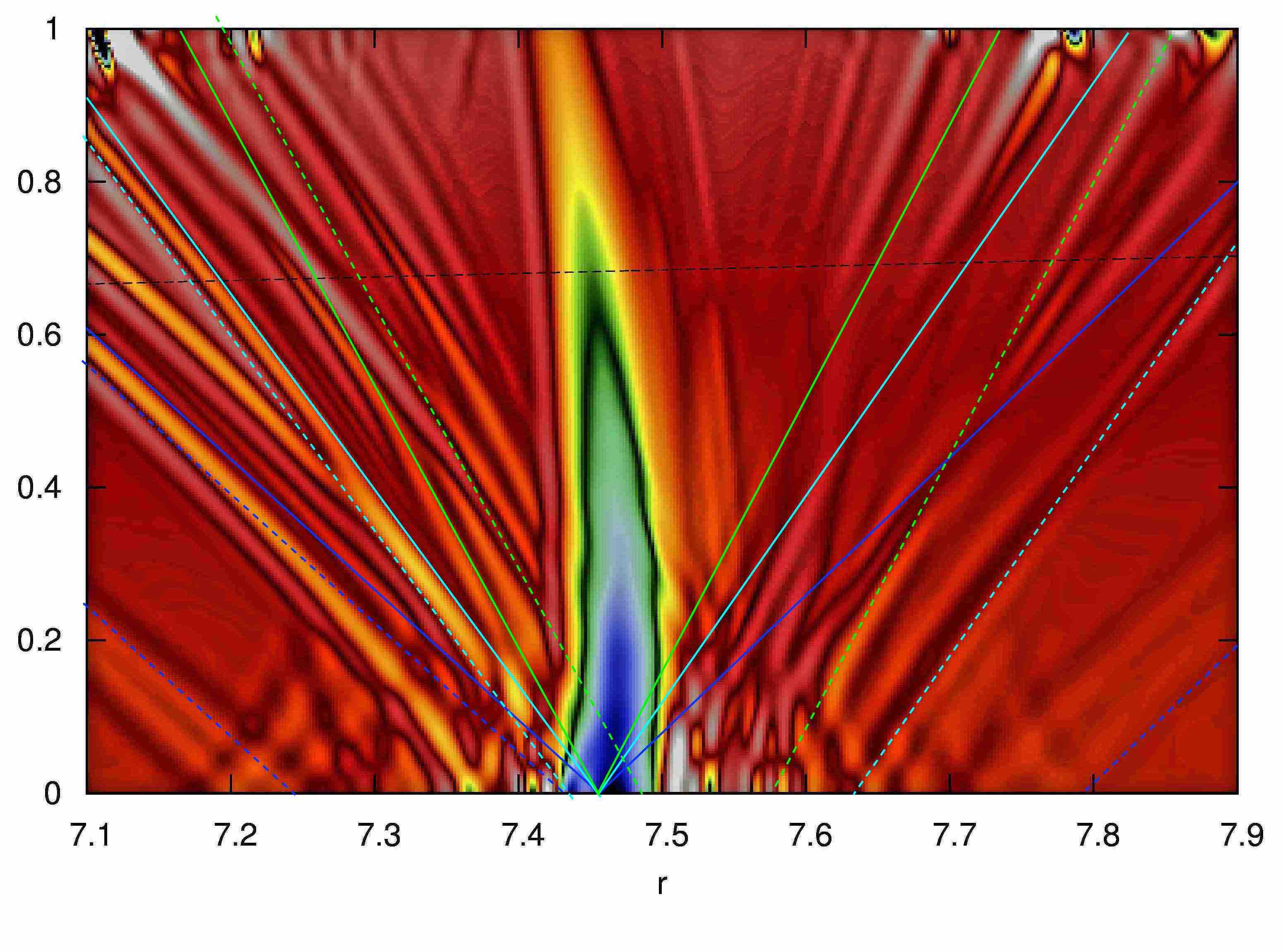}\\
\caption{
Vorticity map after 445 rotations in a $r$-$z$ cross section through the strongest vortex (same plot as in Fig.~\ref{ib3Dplot} but the contrast has been enhanced to make more salient weak features. The horizontal line (black dashed) indicates the height where $N_r^2$ becomes positive. The colored lines (dark-blue, cyan and green) indicate  the position of the baroclinic critical layers, as computed from Eq.~(\ref{slope}) with a source at the center of the mid-plane vortex for $m=16$, $24$ and $32$, respectively. The dashed-colored lines are other critical layer curves with different positions  of the source.} 
\label{zombie-1}
\end{figure*}

We do not see any replication process in the present simulation, but claim that the rays do correspond to  baroclinic critical layers. They clearly resemble the structures shown in \citet{Marcus2013}. The position of the baroclinic critical layers is theoretically given by the relation\
 \begin{equation}
\omega - m\Omega_D(r,z) = \pm {N_z}(r,z)
\end{equation}
where $\omega$ is the frequency of the perturbation, $m$ its azimuthal wavenumber and $N_z$ the Brunt-V\"ais\"al\"a frequency in the vertical direction (see also the recent work of \citet{Umurhan16}).
 With the present vertical structure of the disk and assuming that the mid-plane vortex is the source of the perturbations, we get\\
\begin{equation}
m[\Omega_D(r_v,0) - \Omega_D(r,z)] = \pm \sqrt{\gamma - 1}~ \Omega_K(r) { \frac z{H(r)}}
\label{slope}
\end{equation}
where $r_v$ is the radial position of the mid-plane vortex. 
  For each value of $m$, we obtain two (almost linear) curves in the ($r$-$z$) plane. In Fig.~\ref{zombie-1}, we have plotted 
  these curves for $m= 16, 24, 32$ on the vorticity contours of the perturbation.  We obtain a quite good agreement 
  between the inclinations of the vortex layers and the slopes of the critical layers derived analytically from Eq.~(\ref{slope}). 
   The azimuthal wavenumber of the theoretical critical layer also matches the observed azimuthal structure of the vortex layer: 
   we do observe the formation of two, three and four vortices in the vortex layers associated with the critical layer of 
   azimuthal wavenumber  $m=16$, $24$, and $32$ respectively (see Fig.~\ref{zombie-3D}).
   There is  however an uncertainty on the position of the source which does not seem to be very localized. 
   Several sources between 7.4 to 7.6 seem to contribute.  
   This  is perhaps related to the other vortex which is present in the simulation  (see Fig.~\ref{zombie-3D}) but located at a different radius, outside the cross section plotted in
   Fig.~\ref{zombie-1}.  It then generates a different vortex-layer pattern, which should leave a trace in Fig.~\ref{zombie-1}. 
   This is clearly visible in Fig.~\ref{zombie-3D}: we do observe the formation of a double rows of two vortices and of three vortices corresponding to vortex layers of each mid-plane vortex.    
   This confirms that both the vortex layers and the vortices that form at the boundary of the computation box are not spurious numerical artifacts.
   They are real physical phenomena which deserve further attention.\
   
\begin{figure*}[!h]
\centering
\includegraphics[width=6cm, height=5cm]{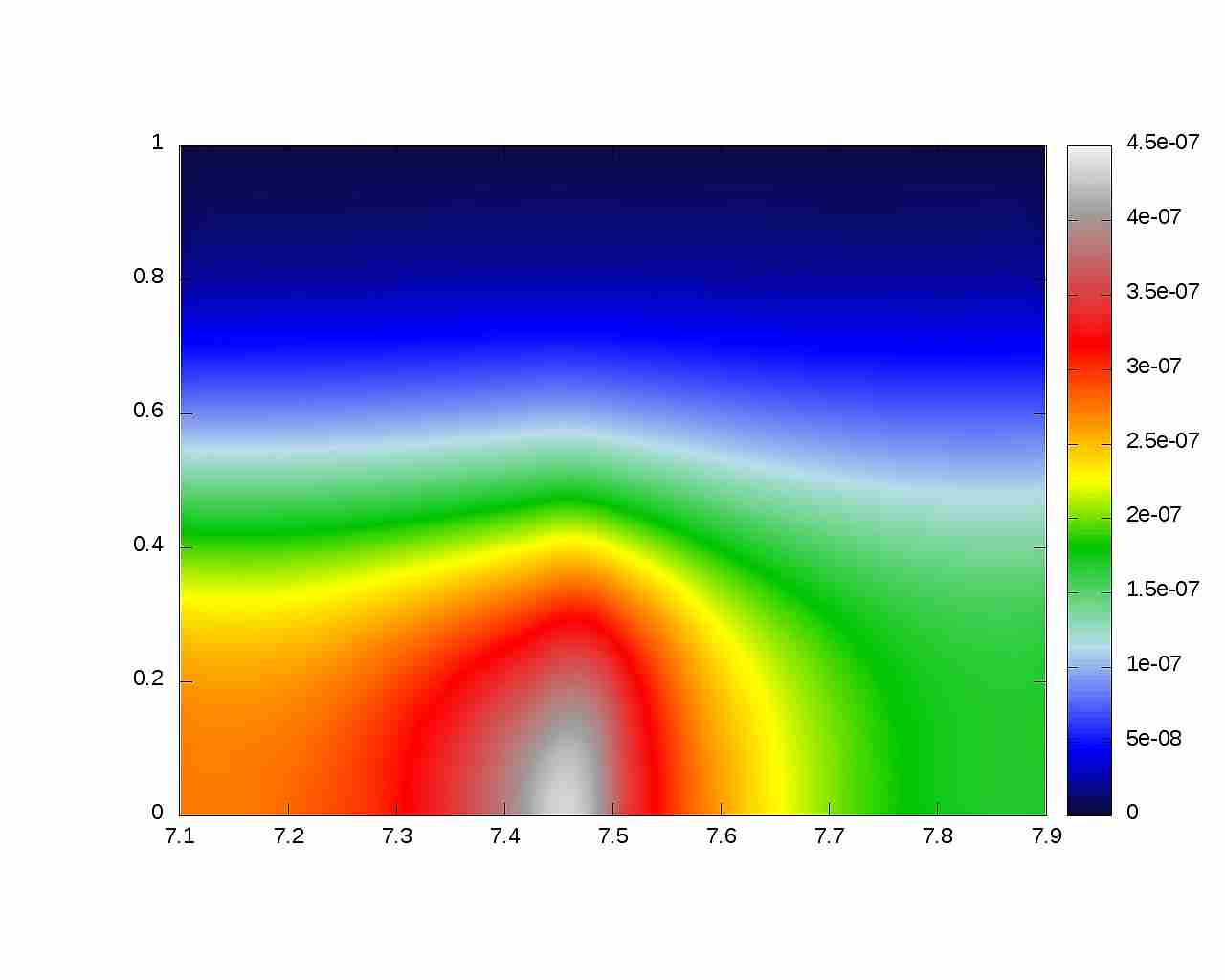}
\includegraphics[width=6cm, height=5cm]{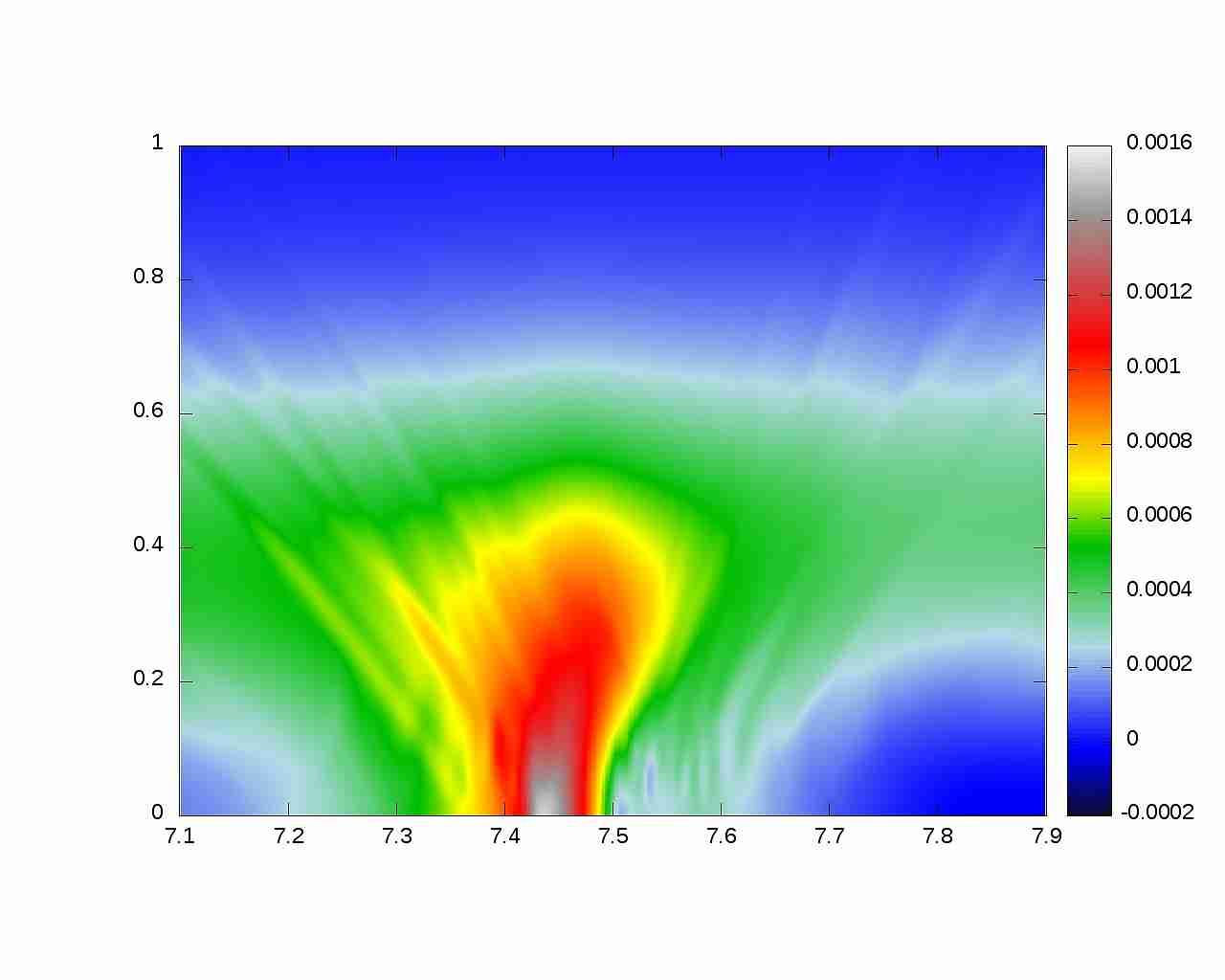}
\includegraphics[width=6cm, height=5cm]{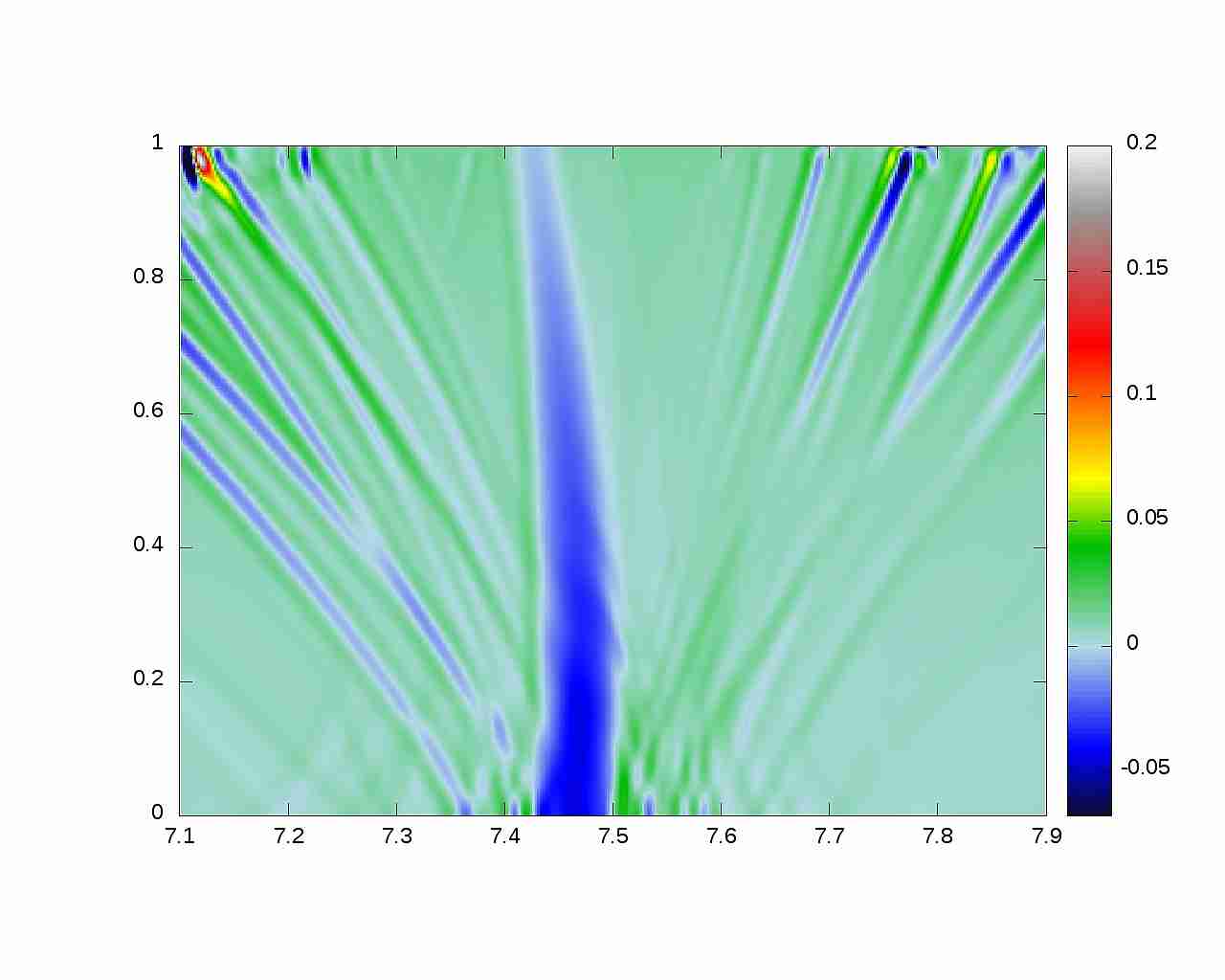}\\
\caption{Maps of the pressure , the density and the divergence of the flow after 445 rotations in a $r$-$z$ cross section through the strongest vortex observed in the simulation. 
$P - \rho_eT(r)$ is plotted on the left, $\rho - \rho_e$ is plotted on the middle and $0.5\ve\nabla\cdot\ve V  - 1/4$ on the right.} 
\label{zombie-2}
\end{figure*}

Before going into deeper studies, it is straightforward to look at other changes in the disk associated with the presence of vortex layers. For example, we have plotted in Fig.~\ref{zombie-2} the distribution of density, pressure and velocity divergence in the same section as in Fig.~\ref{zombie-1}. The most striking feature of this figure is the difference between the plots of  density and pressure with those of velocity divergence and vorticity. The density/pressure contours are mainly limited to the unstable layer of the disk while the vorticity/divergence contours exhibit linear rays that extend through the whole thickness of the disk. 
The rays are also visible on the density contours in the mid-plane layer but they rapidly fade away with increasing amplitude. No rays are by contrast visible on the pressure contours.

\section{Discussion and conclusion}
\label{sec:conc}
In this paper, we have revisited the subcritical baroclinic instability. We have studied a number of unexplored situations, including 2D simulations with heat diffusion and 
3D simulations in a completely stratified disk.\\

In the case of two-dimensional disks, our simulations have confirmed the existence of two main stages in the evolution of the vortices: (i) a first stage during which small vortices form from density bumps seeded in the background, (ii) a second stage during which the small vortices grow and expand by the baroclinic amplification mechanism. 
 We have considered two situations in which thermal transfer is due to either relaxation or heat diffusion. In the case of thermal relaxation, the evolution is found to be the same as described by \citet{Petersen2007a, Petersen2007b}. 
 In the case of heat diffusion, the evolution of the vorticity is more complex and leads to the formation of hollow vortices. These vortices were found to be unstable, but they were not destroyed. Instead, they gave rise to
 coherent vortical structures with a turbulent core that persist for a large number of rotation periods. 
 
In the case of three-dimensional disks, we have first analyzed the influence of the radial distributions of density and temperature on the  stratification properties of the disk. We have shown that the disk could be either unstable in a
mid-plane layer or in a layer distant from the mid-plane. 
We have examined the first situation in which an unstable mid-plane layer is sandwiched between two stable  layers. We have found that, at the beginning of the evolution, the baroclinic instability follows the 
 two-dimensional scenario, i.e. small vortices are formed in the mid-plane and  are then  baroclinically amplified.  However, during their growth, the vortices were found to be stretched out in the vertical direction, leading
 to columnar and long-lived vortex structures extending through the whole width of the disk (including the stable layers). 
 
It would be interesting to examine the second situation where the unstable layer is far from the mid-plane. This situation is a priori possible in the framework of the standard power law model where $q\approx 0.5$ 
and $p>0$ \citep{Andrew2009}. It is possible to imagine that vortices would first develop in the intermediate unstable layer before being advected toward the mid-plane and the upper layer.

The baroclinic instability that we have studied is non-linear in nature.  It can be triggered only in the presence of finite amplitude perturbations. This requires strong enough perturbations of the disk whose origin can only be speculated. It is unlikely that they could be created by the magneto-rotational instability in the vicinity of the dead-zone since this instability tends to inhibit 
vortex formation \citep{Lyra2011}. However, they could be due to another hydrodynamical instability such as the linear version of the baroclinic instability proposed by Klahr \& Hubbard (2014). The vertical shear instability was proposed
by Nelson et al. (2013) as a possible way to trigger the SBI, but recently, Richard et al. (2016) showed that the vertical shear instability leads to the formation of vortices via the Rossby wave instability rather than the SBI.
However, they could be due to another hydrodynamical instability such as the vertical shear instability proposed by \citet{Nelson2013} or a linear version of the baroclinic instability as proposed by \citet{Klahr2014}.

The recent discovery by ALMA of an asymmetry in the dust distribution around Oph IRS 48 \citep{Marel2013, Armitage2013} has been interpreted as an assembly of particles captured by a large-scale Rossby vortex. 
The authors suggested that the vortex had formed by a Rossby wave instability occurring at the edge of the gap opened by an unseen companion orbiting the inner disk. However, it cannot be excluded 
that such a vortex is the result of a baroclinic instability.
\\

Finally, we have also shown that 3D baroclinic vortices are sources of internal vortex layers. These layers  
develop on baroclinic critical layer surfaces and form vortices in the upper-disk region. Whether these vortices could force other vortex layers 
and follow the self-replication mechanism of \citet{Marcus2013_2} remains among the open questions that would be interesting to answer.

\begin{acknowledgements}
Computations were performed on a Bull multicore machine funded by AMU (Aix-Marseille University) and LAM (Laboratoire d'Astrophysique de Marseille); the most demanding computations were performed using HPC ressources from GENCI [TGCC - (grant2013-2014) - t2013046832 - t2014046832 - t2015047407].  3D visualizations were performed with \textit{glnemo2} developed at CeSAM - http://projets.lam.fr/projects/glnemo2. Support by LABEX MEC (ANR-11-LABX-0092) and ANR LIPSTIC (ANR-13-JS05-0004-01) is also acknowledged. We thank an anonymous referee for the review of our manuscript.
\end{acknowledgements}

\bibliographystyle{aa}
\bibliography{Baroclinic}

\end{document}